\documentclass[12pt]{article}

\usepackage{epstopdf,amsfonts,amsmath,amsthm}
\usepackage{longtable}
\usepackage{amssymb,braket,bm}
\usepackage{booktabs,caption,float}
\usepackage{graphicx}
\DeclareGraphicsExtensions{.eps}

\usepackage[textsize=footnotesize]{todonotes}   
\newcommand\encadremath[1]{\vbox{\hrule\hbox{\vrule\kern8pt
\vbox{\kern8pt \hbox{$\displaystyle #1$}\kern8pt}
\kern8pt\vrule}\hrule}}
\def\enca#1{\vbox{\hrule\hbox{
\vrule\kern8pt\vbox{\kern8pt \hbox{$\displaystyle #1$}
\kern8pt} \kern8pt\vrule}\hrule}}

\newcommand\framefig[1]{
\begin{figure}[bth]
\hrule\hbox{\vrule\kern8pt
\vbox{\kern8pt \vbox{
\begin{center}
{#1}
\end{center}
}\kern8pt}
\kern8pt\vrule}\hrule
\end{figure}
}

\newcommand\figureframex[3]{
\begin{figure}[bth]
\hrule\hbox{\vrule\kern8pt
\vbox{\kern8pt \vbox{
\begin{center}
{\mbox{\epsfxsize=#1.truecm\epsfbox{#2}}}
\end{center}
\caption{#3}
}\kern8pt}
\kern8pt\vrule}\hrule
\end{figure}
}
\newcommand\figureframey[3]{
\begin{figure}[bth]
\hrule\hbox{\vrule\kern8pt
\vbox{\kern8pt \vbox{
\begin{center}
{\mbox{\epsfysize=#1.truecm\epsfbox{#2}}}
\end{center}
\caption{#3}
}\kern8pt}
\kern8pt\vrule}\hrule
\end{figure}
}

\renewcommand{\thesection}{\arabic{section}}

\makeatletter
\@addtoreset{equation}{section}
\makeatother
\newtheorem{theorem}{Theorem}[section]

\newtheorem{remark}{Remark}[section]
\newtheorem{proposition}{Proposition}[section]
\newtheorem{lemma}{Lemma}[section]
\newtheorem{corollary}{Corollary}[section]
\newtheorem{definition}{Definition}[section]
\newtheorem{example}{Example}[section]
\def\br{\begin{remark}\rm\small}
\def\er{\end{remark}}
\def\bt{\begin{theorem}}
\def\et{\end{theorem}}
\def\bd{\begin{definition}}
\def\ed{\end{definition}}
\def\bp{\begin{proposition}}
\def\ep{\end{proposition}}
\def\bl{\begin{lemma}}
\def\el{\end{lemma}}
\def\bc{\begin{corollary}}
\def\ec{\end{corollary}}
\def\bex{\begin{example}}
\def\eex{\end{example}}
\def\beaq{\begin{eqnarray}}
\def\eeaq{\end{eqnarray}}

\newcommand{\be}{\begin{equation}}
\newcommand{\ee}{\end{equation}}
\newcommand{\beq}{\begin{equation}}
\newcommand{\eeq}{\end{equation}}
\newcommand{\bea}{\begin{eqnarray}}
\newcommand{\eea}{\end{eqnarray}}

\newcommand{\Tr}{{\operatorname {Tr}}}

\newcommand{\CC}{{\mathbb C}}

\newcommand{\curve}{{\Sigma}}

\newcommand{\modsp}{{\mathcal M}}

\newcommand{\td}{\tilde}

\newcommand{\Sym}{\operatorname{Sym}}
\newcommand{\Asym}{\operatorname{Asym}}
\newcommand{\SL}{{\rm SL}}
\newcommand{\Id}{{\rm Id}}
\newcommand{\Res}{\mathop{\,\rm Res\,}}

\newcommand{\Ai}{{\rm Ai}}
\newcommand{\Bi}{{\rm Bi}}

\newcommand{\uu}{{\bm{u}}}
\newcommand{\xx}{{\bm{x}}}

\newcommand{\Coeff}{D}
\newcommand{\hl}{L}

\usepackage[pdftex]{hyperref}
\hypersetup{colorlinks,urlcolor=magenta,citecolor=red,linkcolor=blue,filecolor=black}
\usepackage[style=alphabetic,maxalphanames=4,giveninits,sorting=nyt]{biblatex}
\renewbibmacro{in:}{}
\usepackage{enumitem}
\usepackage{mathtools}
\begin{document}

\sloppy

\pagestyle{empty}
\addtolength{\baselineskip}{0.20\baselineskip}
\hfill IPhT-T22/128
\begin{center}
\vspace{26pt}
{\large \bf {
A new formula for intersection numbers
}} \\
\vspace{26pt}

{\sl B.\ Eynard}${}^{1,2}$\hspace*{0.05cm}
,
{\sl D.\ Mitsios}${}^{1}$\hspace*{0.05cm}

\vspace{6pt}
${}^{1}$ Université Paris-Saclay, CNRS, CEA, Institut de Physique Théorique, 91191, Gif-sur-Yvette, France.\\
${}^{2}$ CRM, Centre de Recherches Math\'ematiques de Montr\'eal,\\
Universit\'e de Montr\'eal, QC, Canada.
\end{center}
%
%
\vspace{20pt}
\begin{center}
{\bf Abstract}
\end{center}

We propose a new formula to compute Witten--Kontsevich intersection numbers. It is a closed formula, not involving recursion neither solving equations. It only involves sums over partitions of products of factorials, double factorials and Kostka numbers (numbers of semi-standard tableaux of given shape and weight) with bounded weights. As an application, we prove a conjecture of \cite{ELO21} stating that the generating polynomials of the intersection numbers expressed in the basis of elementary symmetric polynomials have an unexpected vanishing of their coefficients.
\pagestyle{plain}
\setcounter{page}{1}

\newpage
\section{Introduction}
\label{sec:intro}

Witten--Kontsevich intersection numbers are extremely useful numbers. They appear in enumerative geometry, in integrable systems, in combinatorics, in random matrix theory, and many areas of geometry and mathematical physics.

They are just rational numbers, as useful as for example Bernoulli numbers. They are usually computed by various recursive algebraic algorithms (Virasoro constraints, cut and join, KdV,...). 
Although they appear in many other areas, their initial definition was motivated by geometry, let us recall how.

\subsection{Witten--Kontsevich intersection numbers}

Witten--Kontsevich intersection numbers were initially introduced in enumerative algebraic geometry, defined as integral of Chern classes on the moduli space of Riemann surfaces.
More precisely, let $(g,n)$ non-negative integers such that $2g-2+n>0$. Let $\modsp_{g,n}=\{ (\curve,p_1,\dots,p_n)\}/\text{Aut}$, the space of Riemann surfaces $\curve$ of genus $g$ with $n$ distinct labelled marked points $p_1,\dots,p_n$, modulo holomorphic automorphisms.
Its Deligne--Mumford compactification, denoted by $\overline{\mathcal M}_{g,n}$, is obtained by adding stable nodal surfaces and makes it into a compact orbifold of complex dimension
\beq
d_{g,n} \coloneqq 3g-3+n.
\eeq
Let $\mathcal L_i\to \overline{\mathcal M}_{g,n}$ the $i$th cotangent line bundle, whose fibre is $T^*_{p_i}\curve$, and let $\psi_i=c_1(\mathcal L_i)$ its 1st Chern class which is a 2-form. If $d_1+\dots+d_n = d_{g,n}$, then $\psi_1^{d_1} \dots \psi_n^{d_n}$ is a volume form, and we define the Witten--Kontsevich intersection number as its integral
\beq
\braket{ \tau_{d_1}\cdots\tau_{d_n} }_g \coloneqq
\braket{\psi_1^{d_1}\cdots\psi_n^{d_n}}_g
\coloneqq \int_{\overline{\mathcal M}_{g,n}} \psi_1^{d_1}\cdots\psi_n^{d_n} \in \mathbb Q.
\eeq
We also define 
\beq
\braket{  \tau_{d_1}\cdots\tau_{d_n} }_g \coloneqq 0 \qquad \text{if} \qquad \sum_{i=1}^n d_i \neq d_{g,n}.
\eeq
These rational numbers are called the Witten--Kontsevich intersection numbers or simply \textbf{intersection numbers} in the context of this article. The notation $\tau_{d_i}=\psi_i^{d_i}$ is called Witten's notation.

The intersection numbers are positive rational numbers.
They play a very important role in many applications in mathematical physics.
Some of them are easy to compute like 
$$
\braket{\tau_0\tau_0\tau_0}_0 =1 
\quad \text{or} \quad 
\braket{\tau_1}_1=\frac{1}{24}.
$$
But for higher $g$ or higher degrees $d_i$ this is a hard task, and the geometric definition is useless for actual computations. An issue is how to compute them in a practical manner?

\bd[Generating polynomials]
For $(g,n)$ such that $n>0$ and $2g-2+n>0$, set
\beq
\begin{split}
A_{g,n}(\uu)
& \coloneqq
\Braket{ \prod_{i=1}^n \frac{1}{1-u_i\psi_i} }_g 
= \sum_{d_1,\dots,d_n}
\braket{  \tau_{d_1}\cdots\tau_{d_n} }_g \prod_{i=1}^n u_i^{d_i} \\
&= \sum_{|\lambda|=d_{g,n}} \braket{\tau_{\lambda_1}\cdots\tau_{\lambda_n}}_g m_\lambda(\uu),
\end{split}
\eeq
where $m_\lambda$ is the monomial symmetric polynomial associated to the partition $\lambda$ (see subsection \ref{subsec:symm:polys} for notation on partitions and symmetric polynomials). For $2g-2+n\leq 0$, set $A_{0,1}(u_1) = u_1^{-2}$ and $A_{0,2}(u_1,u_2) = (u_1+u_2)^{-1}$. Then $A_{g,n}$ is a homogeneous symmetric polynomial of degree
\beq
\deg A_{g,n} = d_{g,n}.
\eeq
We also define the following formal series (filtrated by the degree)
\beq \label{def:An}
A_n=\sum_{g=0}^\infty 2^{g-1} \ A_{g,n}.
\eeq
\ed

The prefactor $2^{g-1}$ is not the most common normalization convention, however it is the one compatible with the standard normalization of Airy function that we use below.

The generating series $A_n(\uu)$ are known in the literature for $n=1,2,3$ (see for instance\footnote{In order to translate the $n$-point function $F$ appearing in \cite{LX11} to our conventions we used homogeneity of $A_{g,n}(\uu)$ to get $A_n(\uu) = 2^{-n/3}F(2^{1/3} \uu)$.} \cite{LX11,ELO21}). They are expressed in terms of the power-sum symmetric polynomials $p_k$ and elementary symmetric polynomials $e_k$.
\begin{itemize}[wide]
\item
$n=1$, due to Witten:
\beq\label{eqn:A1}
    A_1(u) = \frac{e^{\frac{p_3}{12}}}{2}\frac{1}{e_1^2} .
\eeq
\item
$n=2$, due to Dijkgraaf:
\beq\label{eqn:A2}
    A_2(u_1,u_2) = \frac{e^{\frac{p_3}{12}}}{2} \sum_{k=0}^{\infty}
    \frac{1}{(2k+1)!!} e^k_2e^{k-1}_1 .
%
\eeq
\item
$n=3$, due to Zagier (unpublished):
\beq\label{eqn:A3}
A_3(u_1,u_2,u_3)  
= 
\frac{e^{\frac{p_3}{12}}}{2}
\sum_{r,s = 0}^{\infty} 
\frac{r! S_{r}}{2^{r+1} (2r+1)!!}
\frac{\Delta^s}{4^s(r+s+1)!}
\eeq
\noindent
where
\beq
\begin{split}
S_r(u_1,u_2,u_3) 
&=
\frac{(u_1u_2)^r(u_1+u_2)^{r+1}+(u_2u_3)^r(u_2+u_3)^{r+1}+(u_1u_3)^r(u_1+u_3)^{r+1}}{u_1+u_2+u_3} \\
&= 
e_3^r + \sum_{k=0}^r \frac{(-1)^k (r+1)!}{k! (r+1-k)!} \ e_1^{r-k} e_3^k (u_2^{r-k} u_3^{r-k}+u_1^{r-k} u_3^{r-k}+u_1^{r-k} u_2^{r-k}) ,
\\
\Delta(u_1,u_2,u_3) 
&=
(u_1+u_2)(u_2+u_3)(u_1+u_3) 
=
\frac{e_1^3-p_3}{3} 
=
e_1 e_2 - e_3 .
\end{split}
\eeq
\end{itemize}
It is worth mentioning that $S_r$ is a polynomial in $\mathbb{Z}[u_1,u_2,u_3]$.

Closed formulae are also known for fixed genera $g=0$ and $1$, and arbitrary $n$:
\beq
A_{0,n} = e_1^{n-3}
\quad,\quad
 A_{1,n} = \frac{1}{24} \left( e_1^n -\sum_{k=2}^n (k-2)! \ e_k e_1^{n-k} \right).
\eeq
For $g = 2$, $3$ and $4$, see \cite{ELO21}.

\subsection{Main results}

The main result proved in this article is the following theorem, which gives an explicit formula for intersection numbers and their generating functions. A main feature of this formula is that the genus $g$ dependence is encoded in a finite number of $g$-independent coefficients.

\bt[Main theorem] \label{mainTheoremIN}
There exist coefficients $\Coeff_{r,n}(\nu)$, defined in \eqref{eqn:hat:Crn}, depending only on a partition $\nu$ of weight $|\nu|=d_{r,n}$, such that
\beq\label{eqn:int:numbers}
\braket{ \tau_{\lambda_1}\cdots\tau_{\lambda_n} }_g = \frac{1}{24^g}
\sum_{r=0}^{\min(g,\frac{(n-1)(n-2)}{2})}
12^r
\sum_{|\nu|=d_{r,n}} \ \  \sum_{\substack{|\mu|=d_{g,n} \\ \mu\geq\lambda}}
\Coeff_{r,n}(\nu) \, Q_{\nu,\mu} \, \td{K}_{\mu,\lambda} ,
\eeq
or equivalently in terms of generating functions
\beq
A_{g,n}(\uu) =
\frac{1}{24^{g}}
\sum_{r=0}^{\min(g,\frac{(n-1)(n-2)}{2})} 12^{r}
\sum_{|\nu|=d_{r,n}} \ \
\sum_{\substack{ |\mu|=|\lambda|=d_{g,n} \\ \mu \ge \lambda}}
\Coeff_{r,n}(\nu) \, Q_{\nu,\mu} \, \td{K}_{\mu,\lambda} \, m_\lambda(\uu) ,
\eeq
where:
\begin{itemize}
    \item $\td{K}_{\mu,\lambda} \coloneqq N_{\mu,\lambda} K_{\mu,\lambda}$ is the normalized Kostka number, i.e. the Kostka number $K_{\mu,\lambda}$ (counting the number of semi-standard Young tableaux of shape $\mu$ and weight $\lambda$, see \eqref{eqn:Kostka}) multiplied by the combinatorial factor
    \beq\label{eqn:comb:factor}
    N_{\mu,\lambda}
    \coloneqq
    \prod_{i=1}^n
        \frac{\Gamma\left(\mu_i-i+\frac52 \right)\Gamma\left(\frac32 \right)}{\Gamma\left( -i+\frac52 \right)\Gamma\left( \lambda_i+\frac32 \right)}
    =
    2^{|\lambda|} \prod_{i=1}^{n} \frac{\prod_{j=1}^{\mu_i} \left( j-i+\frac32 \right)}{(2\lambda_i+1)!!} .
    \eeq


    \item $Q_{\nu,\mu}$ is given in terms of an inner product involving Schur polynomial $s_{\lambda}$ and the power-sum polynomial $p_{3}$ (see subsection \ref{subsec:symm:polys}):
    \beq
        Q_{\nu,\mu}
        \coloneqq
        \frac{1}{k!}
        \braket{ p_3^{k} s_\nu,s_\mu }
      \quad \text{where } 3k=|\mu|-|\nu|.
    \eeq
    It can be written as a determinant, see \eqref{eq:Qmunuasdet} or appendix \ref{app:Qnumu}.

    \item
    The coefficients $\Coeff_{r,n}(\nu)$ are independent of $g$. The first values are given by
    \beq
    \Coeff_{0,n}(1^{n-3}) = 1
    \qquad 
    \Coeff_{1,4}(2,1,1,0) = \frac12
    \qquad 
    \Coeff_{1,4}(1,1,1,1) = -\frac32
    \qquad 
    \ldots
    \eeq
    Moreover, many of these coefficients vanish, i.e. not all $\nu$ of weight $|\nu| = d_{r,n}$ actually appear.
\end{itemize}
Notice that the sum over $r$ and $\nu$ is independent of the genus $g$ (for $g$ large enough).
\et

In the main body of the text, we will provide equivalent formulations of the above theorem. See subsection \ref{subsec:alternative} for more details.

As an application, we prove a conjecture of \cite{ELO21}, stating that expressing the generating series $A_{g,n}$ in the basis of elementary symmetric polynomials (rather than monomial symmetric or Schur), some simplifications occur.

\bt[{Conjecture of \cite{ELO21}}]
There exist coefficients $C_g(\nu)$ such that
\beq
A_{g,n} = \frac{1}{{24}^g }\sum_{\substack{|\nu|\leq d_{g,n} \\ \nu_i\geq 2, \ \ell(\nu)\leq g}}
C_{g}(\nu) \ e_\nu e_1^{d_{g,n}-|\nu|}
\eeq
and the coefficients $C_g(\nu)$ are independent of $n$, they depend only on the partition $\nu$.
Only partitions $\nu$ of length $\ell(\nu)\leq g$ appear. Here $e_{\nu} \coloneqq \prod_{i=1}^{\ell(\nu)} e_{\nu_i}$ is the product of elementary symmetric polynomials.
\et

\subsection{Previously known algorithms}

Let us recall previously known algorithms to compute intersection numbers.

\begin{itemize}[wide]
\item \textbf{KdV and the Kontsevich matrix model.}
Witten's conjecture \cite{Wit91}, proven by Kontsevich in \cite{Kon92}, states that the generating function $Z$ of intersection numbers, defined as
\beq
\begin{split}
\ln Z(t_1,t_3,t_5,\dots)  
&\coloneqq \sum_{g \ge 0, n>0} \frac{\hbar^{2g-2}}{n!} \ 2^{-(2g-2+n)} \sum_{d_1,\dots,d_n} \braket{\tau_{d_1}\cdots \tau_{d_n}}_g \prod_{i=1}^n (2d_i+1)!! \ t_{2d_i+1} 
\\
&= \sum_{g \ge 0} \left( \frac{\hbar}{2} \right)^{2g-2} \Braket{ e^{\frac12\sum_d \tau_d (2d+1)!! t_{2d+1}} }_g ,
\end{split}
\eeq
is a tau-function of the KdV hierarchy. In proving Witten's conjecture, Kontsevich introduced a formulation of $Z$ as a formal matrix integral (see \cite{Eyn16} for more details):
\beq
Z(t_1,t_3,t_5,\dots) 
= \frac{\prod_{i,j}(\Lambda_i+\Lambda_j)^{\frac12}}{(2\pi\hbar^{\frac13})^{N^2/2}}
\int_{H_N} dM \, e^{\hbar^{-1}\left(\frac13\Tr M^3-\Tr M^2\Lambda\right)} 
\eeq
where  $\Lambda$ is a positive definite $N \times N$  hermitian matrix, $H_N$ denotes the vector space of Hermitian $N \times N$ matrices and
\beq
t_k \coloneqq \hbar \ \Tr \Lambda^{-k} .
\eeq
This matrix integral can be computed explicitly in terms of the Airy function and its derivative:
\beq
Z(t_1,t_3,t_5,\dots) 
= \frac{\hbar^{\frac{N^2-2N}{6}} }{\prod_{i<j}(\Lambda_i-\Lambda_j)}
\det_{1\leq i,j\leq N}\left(\Ai^{(i-1)}(\hbar^{-\frac23}\Lambda_j^2) \right).
\eeq
The KdV equations satisfied by $Z$ are partial differetntial equations with respect to the times $t_1,t_3,t_5,\dots$, that can be turned into a recursive algorithm for computing the itersection numbers. This was the most used computational algorithm after the Witten's conjecture.

\item \textbf{Virasoro constraints and topological recursion.} The intersection numbers satisfy the Virasoro constraints, (which is equivalent to saying that the correlators $W_{g,n}$ defined in \eqref{eqn:corr}, satisfy the topological recursion) :
\beq
\begin{split}
\braket{\tau_{d_1}\cdots\tau_{d_n}}_g
=& \sum_{i=2}^n \frac{(2d_i+2d_1-1)!!}{(2d_1+1)!!(2d_i-1)!!} \langle \tau_{d_1+d_i-1} \prod_{j\neq i} \tau_{d_j} \rangle_{g} \\
& + \frac{1}{2} \sum_{a+b=d_1-2} \frac{(2a+1)!!(2b+1)!!}{(2d_1+1)!!} \braket{\tau_a\tau_b\tau_{d_2}\cdots\tau_{d_n}}_{g-1} \\
& + \frac{1}{2} \sum_{a+b=d_1-2} \sum_{\substack{g_1+g_2=g \\ I_1\sqcup I_2=\{d_2,\dots,d_n\}}} \frac{(2a+1)!!(2b+1)!!}{(2d_1+1)!!} 
\braket{\tau_a\tau_{I_1}}_{g_1} \braket{\tau_b\tau_{I_2}}_{g_2}.
\end{split}
\eeq
Again the Virasoro constraint can be turned into a recursive algorithm to effectively compute intersection numbers.

\item \textbf{Cut-and-join.} In \cite{Ale11}, Alexandrov proved a cut-and-join equation for the generating series of intersection numbers:
\beq
Z = e^{\widehat{W}}1
\eeq
where $\widehat{W}$ is a certain operator in the variables $t_i$ acting on the constant function $1$. The above equation gives a recursive formula in $2g-2+n$ for computing intersection numbers.

\item \textbf{Formulas for the $n$-point functions.}
In the literature, there are several formulas for computing the $n$-point function $A_n$. This includes:
\begin{itemize}
    \item Okounkov's formula \cite{Oko02}, expressing the $n$-point function in terms of $n$-dimensional error-function-type integrals,
    \item Liu--Xu provided a recursive formula for the $n$-point function based on Virasoro constraints \cite{LX11},
    \item determinantal formulas \cite{BDY16,Eyn16}, as discussed in subsection \ref{subsec:det},
    \item Buryak obtained another integral representation $n$-point function formula from the semi-infinite wedge formalism \cite{Bur17}.
\end{itemize}
\end{itemize}

Our main formula will be deduced from the determinantal formula. Compared to the different algorithms, the proposed formula involves only sums over partitions of combinatorial factors. It does not involve integrals, nor solving of KdV equations, no recursion, no differential equations. Moreover, it highlights some unexpected properties of the generating polynomials, like the vanishing of some expansion coefficients conjectured in \cite{ELO21}.

\section{Symmetric polynomials and determinantal formulas}
\label{sec:symm:det}

In this section, we recall some basic facts about symmetric polynomials following \cite{Mac95}, as well as determinantal formulas for intersection numbers \cite{BE09,BDY16,Eyn16}. 

\subsection{Partitions and symmetric polynomials}
\label{subsec:symm:polys}

\subsubsection{Partitions}

Let $\lambda = (\lambda_1,\dots,\lambda_n)$ a partition of a positive integer with $n$ rows $\lambda_1\geq \dots \geq \lambda_n\geq 0$ (we allow empty rows $\lambda_i=0$). We define its weight and its length as
\beq
|\lambda| \coloneqq \sum_{i=1}^n \lambda_i
\qquad , \qquad
\ell(\lambda) \coloneqq \max \{ i \ : \ \lambda_i>0 \} .
\eeq
Its symmetry factor is defined as
\beq\label{def:zlambda}
z_\lambda \coloneqq \prod_{k=0}^{\lambda_1}  (\#\{i \ : \ \lambda_i=k \})!  = (n-\ell(\lambda))! \prod_{k=1}^{\lambda_1}  (\#\{i \ : \ \lambda_i=k \})! \ .
\eeq
The transposed partition given by
\beq
\lambda^T \coloneqq (\lambda'_1,\dots,\lambda'_k)
\qquad , \qquad
\lambda'_i \coloneqq \max \{ j \ : \ \lambda_j\geq i \}.
\eeq
The set of partitions come with a natural partial order, called the dominance order:
\beq
\lambda\geq \mu
\qquad \Leftrightarrow \qquad
\forall i \quad \sum_{j\leq i} (\lambda_j-\mu_j)\geq 0.
\eeq
For a partition with $n$ rows, we define
\beq
\hl_i(\lambda) \coloneqq \lambda_i-i+n .
\eeq
These are positive strictly decreasing numbers $\hl_1(\lambda)>\dots >\hl_n(\lambda)\geq 0$, equal to the hook length of $\lambda$ on the $i$-th row and $1$-st column. 

\subsubsection{Symmetric polynomials}

We collect here some well-known definitions and facts about symmetric polynomials. Here (and in the rest of the paper) all the polynomials will be functions of $n$ variables $\uu = (u_1,\dots,u_n)$.

\paragraph{Elementary symmetric polynomials.} Elementary symmetric polynomials are defined as:
\beq
e_k(\uu) \coloneqq \sum_{1\leq i_1< \dots < i_k\leq n} u_{i_1} \cdots u_{i_k}.
\eeq
We take the convention $e_k=0$ if $k<0$, $e_0=1$, and $e_k=0$ if $k>n$. For a partition $\lambda=(\lambda_1,\dots,\lambda_n)$ of length $\ell(\lambda) \leq n$, completed to $\ell(\lambda)=n$ by adding rows of size $\lambda_i=0$, we set
\beq
e_{\lambda}(\uu) 
\coloneqq \prod_{i=1}^n e_{\lambda_i}(\uu).
\eeq

\paragraph{Monomial and power-sum symmetric polynomials.} Another class of symmetric polynomials is given by the monomial symmetric polynomials: for $\lambda$ as above
\beq\label{defMlambda}
m_\lambda(\uu) \coloneqq \frac{1}{z_\lambda }\sum_{\sigma\in \mathfrak S_n} \prod_{i=1}^n u_i^{\lambda_{\sigma(i)}}.
\eeq
The monomial symmetric polynomial $m_{(k,0,\dots,0)}(\uu)$ is of special interest. It is called the power-sum symmetric polynomial, defined as
\beq
p_k(\uu) \coloneqq \sum_{i=1}^n u_i^k .
\eeq
In the following, we shall mostly use $p_3 = e_1^3-3e_1 e_2 + 3 e_3$.

\paragraph{Schur polynomials.} Another useful basis is given by Schur polynomials. In order to introduce them, define the Vandermonde determinant
\beq
\Delta(\uu) 
\coloneqq \prod_{i<j} (u_i-u_j) = \det u_i^{n-j} = \sum_{\sigma\in \mathfrak S_n} (-1)^\sigma \prod_{i=1}^n u_i^{n-\sigma(i)}.
\eeq
For $\lambda$ as above, the Schur polynomial is defined as
\beq \label{def:Schur}
s_\lambda(\uu)
\coloneqq \frac{1}{\Delta(\uu)} \ \det u_i^{\lambda_j+n-j}
= \frac{1}{\Delta(\uu)} \ \det u_i^{\hl_j(\lambda)} .
\eeq


They can also be expressed as a determinant of complete homogeneous symmetric polynomials:
\beq \label{eqn:Schur:complete:homog}
s_\lambda(\uu) = \det \left( h_{\hl_i(\lambda) - (n-j)}(\uu) \right) ,
\eeq
where
\beq
h_{k}(\uu) \coloneqq \sum_{1 \le i_1 \le \cdots \le i_k \le n} u_{i_1} \cdots u_{i_k} .
\eeq

\paragraph{Schur scalar product.} The space of symmetric polynomials is equipped with a scalar product: if $p$ and $q$ are symmetric polynomials of $n$ variables, we define their Schur scalar product as
\beq\label{defSchurScalarProduct}
\braket{p,q}
\coloneqq
\frac{1}{n!}\Res_{u_i\to 0} \Delta(\uu)p(\uu) \Delta(\uu^{-1})q(\uu^{-1}) \prod_{i=1}^n\frac{du_i}{u_i} 
\eeq
where we set $\uu^{-1} = (u_1^{-1},\dots,u_n^{-1})$. With this scalar product, Schur polynomials form an orthonormal basis:
\beq
\braket{s_\lambda,s_\mu} = \delta_{\lambda,\mu}.
\eeq

\paragraph{Relations.} A useful relation for change of bases is:
\beq \label{eqn:relation:sym:poly}
    \sum_\lambda s_{\lambda^T}(\bm{v})s_\lambda(\bm{u})
    =
    \sum_\lambda e_\lambda(\bm{v})m_\lambda(\bm{u})
    =
    \sum_\lambda m_\lambda(\bm{v})e_\lambda(\bm{u}) .
\eeq 

\paragraph{Kostka numbers.} Schur polynomials can be decomposed on the basis of monomial symmetric polynomials:
\beq\label{eqn:Kostka}
s_\mu = \sum_{\substack{|\lambda|=|\mu| \\ \lambda \le \mu}} K_{\mu,\lambda} \, m_\lambda .
\eeq
The coefficients $K_{\mu,\lambda}$ are called the Kostka numbers, they are non-negative integers counting the number of semi-standard Young tableaux of shape $\mu$ and weight $\lambda$ (see \cite{Kos82,Mac95}). The matrix $K = (K_{\mu,\lambda})$ is upper unitriangular, i.e. $K_{\mu,\lambda} = 0$ unless $\mu \ge \lambda$ and $K_{\mu,\mu} = 1$.

From relation \eqref{eqn:relation:sym:poly}, we deduce the change of basis from elementary symmetric to Schur:
\beq\label{eqn:elementary:to:Schur}
e_{\lambda} = \sum_{\substack{|\mu|=|\lambda| \\ \mu \le \lambda^T}} K_{\mu^T,\lambda} \, s_\mu .
\eeq

Since the matrix $K = (K_{\mu,\nu})$ is upper unitriangular, it is invertible. Denote the elements of the inverse matrix by $K^{-1} = (S_{\lambda,\mu})$, which is again upper unitriangular. From \eqref{eqn:Kostka}, we deduce that $S_{\lambda,\mu}$ are the expansion coefficients of the monomial symmetric polynomials in the basis of Schur polynomials:
\beq
m_\lambda = \sum_{\substack{|\mu|=|\lambda| \\ \mu \le \lambda}} S_{\lambda,\mu} \, s_{\mu} ,
\eeq
and from \eqref{eqn:elementary:to:Schur} we deduce the change of basis from Schur to elementary symmetric:
\beq\label{eqn:Schur:to:elementary}
s_{\mu} = \sum_{\substack{|\lambda|=|\mu| \\ \lambda \ge \mu^T}} S_{\lambda,\mu^T} \, e_\lambda .
\eeq
The coefficients $S_{\lambda,\mu}$ can also be expressed as a determinant:
\beq\label{eqn:S:det}
S_{\lambda,\mu}
=
\sum_{\sigma\in \mathfrak S_n} (-1)^\sigma \det \left(\delta_{\lambda_i+\sigma(i)-1,\mu_j-j+n}\right) .
\eeq

\paragraph{Symmetrization and antisymmetrization.}
In the following, it will be useful to consider the symmetrization and antisymmetrization operators:
\begin{align}
\Sym\bigl[ F(u_1,\dots,u_n) \bigr]
&\coloneqq \sum_{\sigma\in \mathfrak S_n}  F(u_{\sigma(1)},\dots,u_{\sigma(n)} ) , \\
\Asym\bigl[ F(u_1,\dots,u_n) \bigr]
&\coloneqq \sum_{\sigma\in \mathfrak S_n} (-1)^\sigma \, F(u_{\sigma(1)},\dots,u_{\sigma(n)} ) .
\end{align}
With this definition, we have $\Delta(\uu) = \Asym[\prod_{i=1}^n u_i^{n-i}]$. Moreover,
\begin{align}
m_\lambda(\uu)
&= \frac{1}{z_\lambda} \Sym\left[ \prod_{i=1}^n u_i^{\lambda_i} \right] , \\
s_\lambda(\uu) 
&= \frac{1}{\Delta(\uu)} \Asym\left[ \prod_{i=1}^n u_i^{\lambda_i-i+n} \right]
= \frac{1}{\Delta(\uu)} \Asym\left[ \prod_{i=1}^n u_i^{\hl_i(\lambda)} \right] .
\end{align}

\subsection{The Airy function and determinantal formulas}
\label{subsec:det}

Intersection numbers are deeply related to the asymptotic expansion of the Airy function. In particular, they can be generated through the so-called determinantal formulas (see \cite{BE09,BDY16,Eyn16}). Before stating these formulas, let us recall some basic facts about the Airy function.

\subsubsection{The formal Airy function}

\bd[Formal Airy function]
We define the formal ``Airy function" as the following formal series (with exponential prefactor)
\beq\label{def:Airy}
\Ai(x)
\coloneqq
\frac{e^{\frac{2}{3}x^{\frac32}}}{\sqrt{-2}\ x^{\frac14}} \ \sum_{k=0}^\infty
\frac{(6k-1)!!}{2^{3k}3^{2k}(2k)!} \ x^{-3k/2} .
\eeq
It is the asymptotic expansion of the integral (see for instance \cite{BJP15})
\beq
\frac{1}{\sqrt{-\pi}} \ \int_\gamma du \ e^{-\left( \frac{u^3}{3} - u x\right)} \sim_{x\to \infty} \Ai(x).
\eeq
where $\gamma$ is the contour going from $e^{2\pi\mathrm{i}/3}\infty$ to $+\infty$. We define the formal ``Bairy function" by just changing the sign of the square root:
\beq
\Bi(x)
\coloneqq
\frac{e^{-\frac{2}{3}x^{\frac32}}}{\sqrt{-2} \ x^{\frac14}} \ \sum_{k=0}^\infty
(-1)^k \ \frac{(6k-1)!!}{2^{3k}3^{2k}(2k)!} \ x^{-3k/2}.
\eeq
Both formal functions satisfy the linear ODE
\beq\label{eqn:Airy}
\Ai''(x) = x \Ai(x)
\qquad , \qquad
\Bi''(x) = x \Bi(x).
\eeq
Their Wronskian is worth $\Ai(x) \Bi'(x) - \Ai'(x) \Bi(x) = 1$. Indeed, it is easy to see from \eqref{eqn:Airy} that the Wronskian is constant, and we compute it at $x \to \infty$.
\ed

The linear ODE can be transformed into a rank $2$ system as follows.

\bd[Differential system]
Define the $\SL(2,\CC)$ matrix
\beq
\Psi(x) \coloneqq \begin{pmatrix}
\Ai(x) & \Bi(x) \cr
\Ai'(x) & \Bi'(x)
\end{pmatrix}\in \SL(2,\CC) .
\eeq
It is a flat section for the $\SL(2,\CC)$ connection $\nabla \coloneqq d-\mathcal D(x)dx$, i.e. it
satisfies $\nabla \Psi = 0$, i.e. the differential system
\beq
\frac{d}{dx} \Psi(x) = \mathcal D(x) \Psi(x)
\qquad \text{with} \quad
\mathcal D(x) \coloneqq \begin{pmatrix}
0 & 1 \cr x & 0
\end{pmatrix}\in \mathfrak sl(2,\CC).
\eeq
\ed

In the following, we will be interested in the so-called ``adjoint system''.

\bd[Adjoint system]
Define
\beq
M(x) \coloneqq
\Psi(x) \begin{pmatrix}
1 & 0 \cr 0 & 0
\end{pmatrix} \Psi(x)^{-1} 
= \frac12 \Id + \frac12
\Psi(x) \begin{pmatrix}
1 & 0 \cr 0 & -1
\end{pmatrix} \Psi(x)^{-1} 
\eeq
i.e.
\beq
M(x) = \begin{pmatrix}
\Ai(x)\Bi'(x) & -\Ai(x)\Bi(x) \cr \Ai'(x)\Bi'(x) & - \Ai'(x)\Bi(x)
\end{pmatrix}.
\eeq
It is a flat section of the adjoint bundle:
\beq
\frac{d}{dx} M(x) = \left[ \mathcal D(x),M(x) \right].
\eeq
Notice that $M(x) -\frac12 \Id$ is an $\mathfrak sl(2,\CC)$ matrix, and it satisfies the same equation.
\ed

The matrix $M$ is built from a single function $f$, whose properties are given in the following proposition.

\bp
Introduce the formal series
\beq
f(x) \coloneqq -2 \Ai(x)\Bi(x) \in x^{-\frac12}\mathbf Q[[x^{-1}]] .
\eeq
From \eqref{eqn:Airy} it satisfies
\beq
f'''(x) = 4x f'(x) + 2f(x).
\eeq
Its expansion is given by
\beq
f(x) = \frac{1}{\sqrt{x}} \left( 1 + \sum_{k=1}^\infty \frac{(6k-1)!!}{2^{5k}3^k k!} \ x^{-3k}\right)
\eeq
and it can be written as a formal Laplace transform:
\beq
f(x) = \int_0^\infty du \, e^{-xu} \ \frac{1}{\sqrt{\pi u}} e^{\frac{1}{12} u^3} = 
\sum_{k=0}^\infty \int_0^\infty du \, e^{-xu} \ \frac{1}{\sqrt{\pi u}} \ \frac{u^{3k}}{k! 12^k} 
\ .
\eeq
\ep
\begin{proof}
See for instance \cite{BE09,Eyn16}.
\end{proof}

As a consequence:

\bp
The matrix $M$ is expressed as
\beq\label{eqn:M:with:f}
M(x) = \frac12 \Id + \frac12 \begin{pmatrix}
- \frac12 f' & f \cr xf - \frac12 f'' & \frac12 f'
\end{pmatrix}
\eeq
and it can be written as a formal Laplace transform:
\beq\label{eqn:M:Laplace}
M(x) - \frac12 \Id
= -\frac12 \int_0^\infty \frac{du}{\sqrt{\pi u}} e^{-xu} e^{\frac{1}{12}u^3} \
\td M(u)
\quad , \quad
\td M(u) \coloneqq \begin{pmatrix}
-\frac{u}{2} & -1 \cr \frac{u^2}{4}+\frac{1}{2u} &  \frac{u}{2}
\end{pmatrix}.
\eeq
\ep
\begin{proof}
We have
\beq
\frac12 \left(1-\frac12 f' \right) = \frac12(1+\Ai'\Bi + \Ai \Bi') = \frac12(1+2 \Ai \Bi' - (\Ai\Bi'-\Ai'\Bi)) =\Ai \Bi'.
\eeq
Similarly
\beq
\frac12 \left(1+\frac12 f'\right) = \frac12(1-\Ai'\Bi-\Ai \Bi') = \frac12(1-2 \Ai' \Bi - (\Ai\Bi'-\Ai'\Bi)) =-\Ai' \Bi .
\eeq
Then we have
\beq
\begin{split}
xf-\frac12 f''
&= -2 x \Ai \Bi +(\Ai'\Bi+\Ai\Bi')' \\
&= -2 x \Ai \Bi +(2x\Ai\Bi+2\Ai'\Bi') \\
&= 2\Ai'\Bi' .
\end{split}
\eeq
This implies \eqref{eqn:M:with:f}.
Then \eqref{eqn:M:Laplace} is implied by the fact that derivative acts as multiplication by $-u$ in Laplace transform.
\end{proof}

\subsubsection{Determinantal formulas for intersection numbers}

In the introduction we defined the generating polynomials $A_{g,n}$ for the intersection numbers. There are other ways to encode them into a generating function. Among them, the so-called ``correlators" plays an important role.

\bd[Correlators]
For $(g,n)$ such that $n>0$ and $2g-2+n>0$, set
\beq\label{eqn:corr}
W_{g,n}(\xx)
\coloneqq (-2)^{-(2g-2+n)}
\sum_{d_1,\dots,d_n}
\braket{  \tau_{d_1}\cdots\tau_{d_n} }_g \prod_{i=1}^n \frac{(2d_i+1)!! dx_i}{2 \, x_i^{d_i+\frac32}} .
\eeq
$W_{g,n}$ is a symmetric $\otimes^n$ differential-form. For $2g-2+n\leq 0$, set
\beq
\begin{aligned}
W_{0,1}(x) &\coloneqq \sqrt{x} \, dx , \\
W_{0,2}(x_1,x_2) &\coloneqq \frac{1}{4\sqrt{x_1 x_2} \ (\sqrt{x_1}-\sqrt{x_2})^2} \, dx_1\otimes dx_2.
\end{aligned}
\eeq
We also define the following formal series (filtrated by the degree)
\beq
W_n \coloneqq \sum_{g=0}^{\infty} W_{g,n} .
\eeq
\ed

The forms $W_{g,n}$ are those that satisfy topological recursion \cite{EO07}, they originated from the Kontsevich matrix model \cite{Kon92}, and from the Strebel graphs combinatorial decomposition of $\mathcal M_{g,n}$ (see for example \cite{Eyn16}).

Notice that $W_{g,n}$ and $A_{g,n}$ are related by Laplace transforms.

\bl
\label{lemmaWALaplace}
If $2g-2+n>0$, we have
\beq
W_{g,n}(\xx)
=
(-1)^n \ 2^{g-1} \ \prod_{i=1}^n dx_i
\int_{[0,\infty[^n} \prod_{i=1}^n \frac{\sqrt{u_i}du_i}{\sqrt\pi}e^{-u_i x_i}
\ A_{g,n}(\uu)
\eeq
where it is assumed that all $x_i$'s have positive real part $\Re x_i>0$.
\el
\begin{proof}
From the exponential integral
\beq
\int_0^\infty \frac{\sqrt{u} \ du}{\sqrt{\pi}} e^{-xu} u^d
= \frac{(2d+1)!!}{2^{d+1} \, x^{d+\frac32}}
\eeq
we deduce
\beq
\begin{split}
\prod_{i=1}^n dx_i
\int_{[0,\infty[^n} \prod_{i=1}^n \frac{\sqrt{u_i}du_i}{\sqrt\pi}e^{-u_i x_i}
\ A_{g,n}(\uu)
&= \frac{1}{2^{3g-3+n}} (-2)^{2g-2+n} \ W_{g,n}(\xx) \\
&= (-1)^n \ 2^{-(g-1)} \ W_{g,n}(\xx) .
\end{split}
\eeq
\end{proof}

We can finally state the determinantal formula for the intersection numbers.

\bt[Determinantal formulas \cite{BE09,BDY16,Eyn16}]\label{Thdetformulas}
For $n \ge 3$, the correlators are given by
\beq
W_n(\xx)
= \sum_{\sigma \in \mathfrak S_n^{\textup{cycl}}} (-1)^\sigma
\frac{\prod_{i=1}^n dx_i}{\prod_{i=1}^n (x_i-x_{\sigma(i)})}
\Tr \left( \prod_{i=1}^n M(x_{\sigma^i(1)}) \right)
\eeq
where $\mathfrak S_n^{\textup{cycl}}$ is the set of permutations that have only one cycle, also called ``cyclic permutations''.
\et

For $n =1$ and $2$, we refer to loc. cit. for the appropriate formula. Since these cases are completely understood (see \eqref{eqn:A1} and \eqref{eqn:A2}), we will only focus on $n \ge 3$.

Notice that permutations with a single cycle have signature $(-1)^\sigma = (-1)^{n-1}$.
This implies the following statement.

\bp
For $n \ge 3$, the correlators are given by
\beq\label{eqWgnSymTRM}
W_n(\xx)
=
\frac{(-1)^{n-1}}{n} \Sym \left[ \frac{\prod_{i=1}^n dx_i}{\prod_{i=1}^n (x_i-x_{i+1})}
\Tr \left( \prod_{i=1}^n M(x_{i}) \right) \right]
\eeq
where we take the convention $x_{n+i}=x_i$.
Moreover we can get rid of the identity in the trace:
\beq
W_n(\xx)
=
\frac{(-1)^{n-1}}{n} \Sym \left[ \frac{\prod_{i=1}^n dx_i}{\prod_{i=1}^n (x_i-x_{i+1})}
\Tr \left( \prod_{i=1}^n \left(M(x_{i})-\frac12 \Id\right) \right) \right] .
\eeq
\ep
\begin{proof}
\eqref{eqWgnSymTRM} is obvious. The proof that the identity can be subtracted is given in appendix \ref{ApplemmaTraceId}. It is a general property in Lie algebras theory: adding an element of the center leaves the formula invariant.
\end{proof}

\section{Proof of the main results}
\label{sec:proof}

The goal of this section is to prove the main formula from theorem \ref{mainTheoremIN}. The central idea is to express $W_{g,n}$ in two ways: on the one hand, as the Laplace of $A_{g,n}$, and on the other hand in terms of the determinantal formula involving the matrix $M$. The matrix $M$, in its turn, is expressed as a Laplace transform of $e^{u^3/12} \td M$. A manipulation of these identities gives the final result. At a more technical level, the Laplace transform of the determinantal formula will correspond to a certain operator $H$, that we now introduce, acting on the space of symmetric functions.

In this section, we assume $n \ge 3$. The cases $n = 1$, $2$ are fully understood, and can be considered separately.

\subsection{The operator \texorpdfstring{$H$}{H}}
\label{subsec:H}

The operator $H$ we are about to introduce will be crucial in the proof of the main formula. It is an endomorphism in the space of symmetric polynomials in $n$ variables.

\bd[Operator $H$]\label{defH}
Let the operator $H$ acting on symmetric functions $f(\uu)$ of $n$ variables be defined by
\beq\label{eqdefH}
H(f(\uu)) \coloneqq \frac{e_n^{n-\frac32}}{\Delta(\uu) D_n} \ \Delta(d/d\uu) \left( \sqrt{e_n} \ f(\uu) \right)
\eeq
where $\Delta(d/d\uu)$ is the differential operator
\beq
\Delta(d/d\uu) \coloneqq \prod_{i<j} \left(\frac{d}{du_i} - \frac{d}{du_j} \right)
\eeq
and $D_n$ is the combinatorial factor
\beq
D_n \coloneqq \frac{G(n-\frac12)}{G(\frac12) \Gamma(-\frac12)^{n-1}}
= (-1)^{n-1} 2^{-\frac{n(n-1)}{2}} \prod_{k=1}^{n-2} (2k-1)!! .
\eeq
Here $G$ is the Barnes G-function\footnote{
    A defining property of the Barnes $G$-function is the recursion $G(x+1) = \Gamma(x) G(x)$ with the initial condition $G(1) = 1$. From the defining property, it is easy to deduce
    \[
        \frac{G(n-\frac{1}{2})}{G(\frac{1}{2})}
        =
        \Gamma(\tfrac{1}{2})^{n-1} 2^{-\frac{(n-1)(n-2)}{2}} \prod_{k=1}^{n-2} (2k-1)!! .
    \]
    Dividing by $\Gamma(-\frac12)^{n-1}$ yields the above relation.
}.
\ed

We collect here some properties of the operator $H$ that will be useful in the proof of the main formula. See appendix \ref{app:H} for a proof.

\bl[Properties of the operator $H$]\label{lemma:H}
~
\begin{enumerate}
    \item\label{H:end:deg}
    The operator $H$ is an endomorphism in the space of symmetric polynomials in $n$ variables. Moreover for homogeneous polynomials, it conserves the degree.

    \item\label{H:e1}
    $H(1) = 1$ and
    \beq
    H(e_1^k f) = e_1^k H(f).
    \eeq

    \item\label{H:monomial:Schur}
    The operator $H$ sends monomial symmetric polynomials to linear combinations of Schur polynomials:
    \beq
    H(m_\lambda) = \sum_{\substack{|\mu|=|\lambda| \\ \mu\leq\lambda}} \td{S}_{\lambda,\mu} \, s_\mu .
    \eeq
    Conversely, $H^{-1}$ sends Schur polynomials to linear combinations of monomial symmetric polynomials:
    \beq
    H^{-1}(s_\mu) = \sum_{\substack{|\lambda|=|\mu| \\ \lambda\leq\mu}} \td{K}_{\mu,\lambda} \, m_\lambda .
    \eeq
    Here $\td{K}_{\mu,\lambda} \coloneqq N_{\mu,\lambda} K_{\mu,\lambda}$ are the normalized Kostka numbers, and $\td{S}_{\lambda,\mu} = S_{\lambda,\mu}/N_{\mu,\lambda}$ are the matrix elements of the inverse matrix.

    \item\label{H:elementary}
    $H$ is a triangular operator in the basis of elementary symmetric polynomials:
    \beq
    H(e_\lambda) = \sum_{\mu \geq \lambda} H_{\lambda,\mu} \, e_\mu,
    \eeq
    where the coefficients $H_{\lambda,\mu}$ are independent of $n$ and explicitly given by
    \beq
    H_{\lambda,\mu}
    =
    \sum_{\mu^T \le \rho \le \sigma \le \tau \le \lambda^T}
    K_{\tau^T,\lambda} K_{\tau,\sigma} \td{S}_{\sigma, \rho} S_{\mu,\rho^T} .
    \eeq
    Moreover, if $\lambda$ is a hook (one row and one column) then:
    \beq\label{eqn:H:hook}
    H(e_k e_1^l)  =  \frac{1}{N_{(1)^k,(1)^k}} \, e_k e_1^l = (-1)^k \frac{3^{k-1}}{(2k-5)!!}  \, e_k e_1^l .
    \eeq
\end{enumerate}
\el

\subsection{Proof of the main formula}
\label{subsec:main:formula}

In this subsection we study the action of $H$ on the generating series $A_n$ of intersection numbers, normalized by $\tfrac{1}{2}e^{\frac{p_3}{12}}$. This prefactor is a common in all generating series, see \eqref{eqn:A1}--\eqref{eqn:A3}.

\bd
Define the symmetric function $P_n$ by setting
\beq
H(A_n)
\eqqcolon
\frac{e^{\frac{p_3}{12}}}{2} P_n .
\eeq
\ed

Thanks to the determinantal formulas, we can express $P_n$ in terms of the matrix $\td M$ defined in \eqref{eqn:M:Laplace}.

\bp
$P_n$ is given by
\beq\label{eqn:hatP}
P_n(\uu)
=
\frac{e_n^{n-\frac32} }{\Delta(\uu) D_n}
\frac{ e^{ -\frac{p_3}{12} }}{n 2^{n-1}} 
\Asym\left[ 
\prod_{i=1}^{n-2}\prod_{j=i+2}^{n} \left(\frac{d}{du_i}-\frac{d}{du_j}\right) 
\left(
\frac{ e^{ \frac{p_3}{12} } }{\sqrt{e_n}} \ \Tr \prod_{i=1}^n \td M(u_i)
\right)
\right].
\eeq
\ep
\begin{proof}
Starting from the determinantal formula (theorem \ref{Thdetformulas}) multiplied by the Vandermonde, we have 
\beq
\begin{split}
& n \Delta(\xx) \frac{W_n(\xx)}{\prod_{i=1}^n dx_i} \\
& = (-1)^{n-1} \ \Asym \left[ 
\frac{\Delta(\xx)}{\prod_{i=1}^n (x_i-x_{i+1})}
\Tr \left( \prod_{i=1}^n M(x_{i}) \right) \right] \\
& = -\frac{1}{2^n} \ \Asym \left[ 
  \prod_{i=1}^{n-2}\prod_{j=i+2}^{n} (x_i-x_j)
\int \prod_{i=1}^n \frac{du_i}{\sqrt{\pi u_i}} e^{-u_i x_i} e^{\frac{u_i^3}{12}}  \Tr \left( \prod_{i=1}^n \td M(u_{i}) \right) \right] \\
& = -\frac{1}{2^n} \ \Asym \left[ 
\int \prod_{i=1}^n \frac{du_i}{\sqrt{\pi u_i}} e^{\frac{u_i^3}{12}} \Tr \left( \prod_{i=1}^n \td M(u_{i}) \right) \times \right.
\\
&\hspace{4.5cm} \left. \times
\prod_{i=1}^{n-2}\prod_{j=i+2}^{n} \left(-\frac{d}{du_i}+\frac{d}{du_j}\right) \prod_{i=1}^n e^{-u_i x_i} 
\right] \\
& = \frac{(-1)^{n}}{2^n} \ \Asym \left[ 
\int \prod_{i=1}^n \frac{du_i}{\sqrt\pi} e^{-u_i x_i}
\prod_{i=1}^{n-2}\prod_{j=i+2}^{n}  \left(\frac{d}{du_i}-\frac{d}{du_j}\right) 
\frac{e^{\frac{p_3}{12}}}{\sqrt{e_n}} \Tr \left( \prod_{i=1}^n \td M(u_{i}) \right) 
\right] . \\
\end{split}
\eeq
The last equality follows by integration by parts. On the other hand, using the Laplace transform (lemma \ref{lemmaWALaplace}) the LHS is worth
\beq
\begin{split}
n \Delta(\xx) \frac{W_n(\xx)}{\prod_{i=1}^n dx_i} 
&= (-1)^n  \Delta(\xx)
\int \prod_{i=1}^n \frac{du_i }{\sqrt\pi} e^{-u_i x_i} \sqrt{e_n} A_{n}(\uu) \\
& =  n (-1)^n 
\int \prod_{i=1}^n  \frac{du_i }{\sqrt\pi} \sqrt{e_n} A_{n}(\uu) \ \Delta(-d/d\uu) \prod_{i=1}^n e^{-u_i x_i} \\
& = n (-1)^n
\int \prod_{i=1}^n \frac{du_i }{\sqrt\pi} e^{-u_i x_i} \Delta(d/d\uu) \left( \sqrt{e_n} A_{n}(\uu) \right) . \\
\end{split}
\eeq
Again, the last equality follows by integration by parts. This shows that
\beq
n 2^{n} \Delta(d/d\uu) \left(\sqrt{e_n} A_{n}\right)
= 
\Asym \left[ 
  \prod_{i=1}^{n-2}\prod_{j=i+2}^{n}  \left(\frac{d}{du_i}-\frac{d}{du_j}\right) 
 \frac{e^{\frac{p_3}{12}}}{\sqrt{e_n}} \Tr \left( \prod_{i=1}^n \td M(u_{i}) \right) 
\right] .
\eeq
The LHS is recognized as the operator $H$ of Definition \ref{defH}, up to normalization.
\end{proof}

From the above expression, we can deduce that $P_n$ (for $n \ge 3$) is a symmetric polynomial of degree $3 \frac{(n-1)(n-2)}{2} - 3 + n$, with homogeneous components of degree $d_{r,n}$. In other words, the degree of the homogeneous components jumps by $3$.

\bp
For $n \ge 3$, $P_n$ is a symmetric polynomial degree $3 \frac{(n-1)(n-2)}{2} - 3 + n$. Moreover, its homogeneous components have degree $d_{r,n}$ for $r = 0,\dots,\frac{(n-1)(n-2)}{2}$. We will denote them by $P_{r,n}$. In other words,
\beq \label{eq:hatPn:sum:of:hatPrn}
P_n = \sum_{r=0}^{\frac{(n-1)(n-2)}{2}} P_{r,n}
\qquad , \qquad
\deg P_{r,n} = d_{r,n} .
\eeq
\ep
\begin{proof}
We first prove that $P_n$ is a symmetric polynomial. Let us analyze the operations in \eqref{eqn:hatP} step-by-step.

\begin{itemize}[wide]
    \item
    Notice that the exponential $e^{\frac{p_3}{12}}$ cancels out.

    \item
    We now take derivatives of half integer powers of the $u_i$'s, and then multiply by $e_n^{n-3/2}$. Therefore the result has only integer powers (positive or negative) and is a rational function of the $u_i$'s, with possible poles at $u_i=0$.

    \item
    We then divide an antisymmetric function by the Vandermonde, so the result is a symmetric function.

    \item
    The lowest possible power of $u_i$ is given by: $u_i^{-1}$ coming from $\td M(u_i)$, times $u_i^{-1/2}$ from $1/\sqrt{e_n}$, times $u_i^{-(n-3)}$ from the derivatives. This gives a total power of $u_i^{-(n-3/2)}$ which is compensated by $e_n^{n-3/2}$. Thus, each $u_i$ has positive integer powers, i.e. the result is a symmetric polynomial.
\end{itemize}
Let us now compute the degree of $P_n$. The highest possible homogeneous degree in the trace of powers of the $\td M(u_i)$ is $n$. After multiplying by $1/\sqrt{e_n}$, the highest possible homogeneous degree is thus $n/2$. We then take $n(n-1)/2-n $ derivatives after multiplying by the exponentials, and each cubic exponential can yield at most $u_i^2$. Therefore the degree is at most:
\beq
2\left( \frac{n(n-1)}{2}-n \right) + \frac{n}{2} . 
\eeq
Then multiplying by $e_n^{n-3/2}$ and dividing by the Vandermonde gives at most
\beq \label{maxdeghatPn}
2\left( \frac{n(n-1)}{2}-n \right) + \frac{n}{2} + n\left(n-\frac{3}{2} \right) - \frac{n(n-1)}{2}
=
3 \frac{(n-1)(n-2)}{2} -3 + n .
\eeq
To conclude the proof, we simply have to prove that the homogeneous components $P_{r,n}$ have degree $d_{r,n}$, i.e. they jump by 3 in degree. From the definition of $P_{n}$ and the decomposition of $A_n$ into homogeneous components, we see that
\beq\label{eq:hatPnasHAnexp}
\frac12 P_n = e^{-\frac{p_3}{12}} H(A_n) = \sum_{g\geq 0}\sum_{k\geq 0} 2^{g-1}H(A_{g,n})  \frac{(-1)^k}{12^k k!} p_3^k .
\eeq
The operator $H$ conserves the degree (see lemma \ref{lemma:H}, property (\ref{H:end:deg})), which implies that the right hand side can have degrees $3g-3+n+3k$, i.e. $n-3$ plus multiples of 3, i.e. of the form $d_{r,n} = 3r-3+n$. Therefore we can decompose $P_n$ into homogeneous terms whose degree are $3r-3+n$:
\beq
P_n = \sum_{r \ge 0} P_{r,n} \qquad \deg P_{r,n} = d_{r,n}.
\eeq
The maximum degree \eqref{maxdeghatPn} then implies that  $r\leq r_{\max}=(n-1)(n-2)/2$.
\end{proof}
%


The coefficients of the decomposition of $P_{r,n}$ in the basis of Schur polynomials will play an important role in the main result of this paper. 

\bd[Decomposition on Schur's]\label{def:hatCrn}
We decompose the homogeneous symmetric polynomial $P_{r,n}$ onto the basis of Schur polynomials
\beq\label{eqn:hat:Crn}
P_{r,n} = \sum_{|\nu|=d_{r,n}} \Coeff_{r,n}(\nu) \, s_\nu.
\eeq
\ed

The decomposition of $P_{n}$ in the Schur basis for $n=3$, $4$ and $5$ is displayed in Table \ref{table:hatP:Schur}. Observe that very few partitions of size $|\nu|=d_{r,n}$ actually appear: many coefficients $\Coeff_{r,n}(\nu)$ do in fact vanish. This unexpected vanishing will be analyzed in details in the next section (in the basis of elementary symmetric polynomials).

\begin{table}[ht]
\centering
\begin{tabular}{r | l}
\toprule 
    $\bm{n=3}$, $\bm{r=0}$ &
    $P_{0,3}=s_{\emptyset}$ \\[.5ex]
    $\bm{1}$ &
    $P_{1,3}=\tfrac{1}{2} s_{(1,1,1)}$ \\[.5ex]
\midrule
    $\bm{n=4}$, $\bm{r=0}$ &
    $P_{0,4}=s_{(1)}$ \\[.5ex]
    $\bm{1}$ &
    $P_{1,4}=\tfrac{1}{2} s_{(2,1,1)}-s_{(1,1,1,1)}$ \\[.5ex]
    $\bm{2}$ &
    $P_{2,4}=-\tfrac{1}{6} s_{(2,2,2,1)}+\tfrac1{12}s_{(3,2,1,1)}$ \\[.5ex]
    $\bm{3}$ &
    $P_{3,4}=\tfrac{1}{24} s_{(3,3,2,2)}$ \\[.5ex]
\midrule
    $\bm{n=5}$, $\bm{r=0}$ &
    $P_{0,5} = s_{(2)}+s_{(1,1)}$ \\[.5ex]
    $\bm{1}$ &
    $P_{1,5}=
    \tfrac{1}{2} s_{(3,1,1)} 
    + \tfrac{1}{2} s_{(2,2,1)} 
    - \tfrac{1}{2} s_{(2,1,1,1)} 
    + \tfrac{17}{10} s_{(1,1,1,1,1)}$ \\[.5ex]
    $\bm{2}$ &
    $P_{2,5}=
    \tfrac{1}{12} s_{(4,2,1,1)} 
    + \tfrac{1}{10} s_{(4,1,1,1,1)} 
    + \tfrac{1}{12} s_{(3,3,1,1)} 
    - \tfrac{11}{60} s_{(3,2,1,1,1)} 
    - \tfrac{1}{12} s_{(3,2,2,1)}$ \\[.5ex]
    & $\qquad\quad
    - \tfrac{1}{6} s_{(2,2,2,2)} 
    + \tfrac{2}{3} s_{(2,2,2,1,1)}$ \\[.5ex]
    $\bm{3}$ &
    $P_{3,5}=
    \tfrac{1}{120} s_{(5,3,1,1,1)} 
    + \tfrac{1}{120} s_{(5,2,2,1,1)} 
    + \tfrac{1}{120} s_{(4,4,1,1,1)} 
    + \tfrac{1}{24} s_{(4,3,2,2)}$ \\[.5ex]
    & $\qquad\quad
    - \tfrac{1}{40} s_{(4,3,2,1,1)} 
    + \tfrac{1}{60} s_{(4,2,2,2,1)} 
    + \tfrac{1}{24} s_{(3,3,3,2)} 
    + \tfrac{1}{30} s_{(3,3,3,1,1)} $ \\[.5ex]
    $\bm{4}$ &
    $P_{4,5}=
    \tfrac{1}{240} s_{(5,4,2,2,1)} 
    + \tfrac{1}{240} s_{(5,3,3,2,1)} 
    - \tfrac{1}{240} s_{(5,3,2,2,2)} 
    - \tfrac{1}{240} s_{(4,4,3,2,1)} $ \\[.5ex]
    & $\qquad\quad
    - \tfrac{11}{720} s_{(4,3,3,2,2)} 
    - \tfrac{1}{240} s_{(4,3,3,3,1)} 
    + \tfrac{1}{60} s_{(3,3,3,3,2)} $ \\[.5ex]
    $\bm{5}$ &
    $P_{5,5}=
    \tfrac{1}{1440} s_{(5,5,3,2,2)} 
    - \tfrac{1}{1440} s_{(5,4,3,3,2)} 
    + \tfrac{1}{1440} s_{(5,4,4,2,2)} 
    - \tfrac{1}{720} s_{(5,3,3,3,3)} $ \\[.5ex]
    & $\qquad\quad
    - \tfrac{11}{360} s_{(4,4,3,3,3)} 
    - \tfrac{1}{1440} s_{(4,4,4,3,3)} $ \\[.5ex]
    $\bm{6}$ &
    $P_{6,5}=
    \tfrac{1}{2880} s_{(5,5,4,3,3)} 
    + \tfrac{1}{2880} s_{(5,4,4,4,3)} 
    - \tfrac{1}{960} s_{(4,4,4,4,4)} $ \\[.5ex]
\bottomrule
\end{tabular}
\caption{The homogeneous components of $P_n$ for $n = 3, 4, 5$ in the Schur basis.}
\label{table:hatP:Schur}
\end{table}

We are now ready to prove the main result, theorem \ref{mainTheoremIN}, by explicitly computing the action of the operator $H^{-1}$.

\bt[Main theorem]
The following formula holds:
\beq\label{eq:main:formula:tau}
\braket{ \tau_{\lambda_1}\cdots\tau_{\lambda_n} }_g = \frac{1}{24^g}
\sum_{r=0}^{\min(g,\frac{(n-1)(n-2)}{2})}
12^r
\sum_{|\nu|=d_{r,n}} \ \ \sum_{\substack{|\mu|=d_{g,n} \\ \mu\geq\lambda}}
\Coeff_{r,n}(\nu) \, Q_{\nu,\mu} \, \td{K}_{\mu,\lambda} ,
\eeq
or equivalently in terms of generating functions
\beq\label{eq:main:formula:Agn}
A_{g,n}(\uu) =
\frac{1}{24^{g}}
\sum_{r=0}^{\min(g,\frac{(n-1)(n-2)}{2})} 12^{r}
\sum_{|\nu|=d_{r,n}} \ \
\sum_{\substack{ |\mu|=|\lambda|=d_{g,n} \\ \mu \ge \lambda}}
\Coeff_{r,n}(\nu) \, Q_{\nu,\mu} \, \td{K}_{\mu,\lambda} \, m_\lambda(\uu) .
\eeq
Here $\Coeff_{r,n}(\nu)$ are defined in \eqref{eqn:hat:Crn}, $\td{K}_{\mu,\lambda}$ are the normalized Kostka numbers, and $Q_{\nu,\mu}$ is the inner product $ Q_{\nu,\mu} =\frac{1}{k!} \left< p_3^{k} s_\nu,s_\mu \right>$ where $3k=|\mu|-|\nu|$.
\et
\begin{proof}
Applying $H^{-1}$ to the equality $H(A_n) = \frac{e^{\frac{p_3}{2}}}{2}P_n$ and decomposing it into homogeneous components, we find
\beq\label{eq:AgnhatP}
A_{g,n} = \frac{1}{24^g} \sum_{r=0}^{\min(g,\frac{(n-1)(n-2)}{2})} \frac{12^r }{(g-r)!} H^{-1}(p_3^{g-r} P_{r,n}) .
\eeq
From the decomposition of $p_3^{g-r}P_{r,n}$ on the basis of Schur polynomials, i.e.
\beq
p_3^{g-r}P_{r,n} = \sum_{|\mu|=d_{g,n}} \left< s_\mu,p_3^{g-r} P_{r,n}\right> s_\mu,
\eeq
we get
\beq
A_{g,n}
=
\frac{1}{24^g} \sum_{|\mu|=d_{g,n}} H^{-1}(s_\mu) \sum_{r=0}^{\min(g,\frac{(n-1)(n-2)}{2})} \frac{12^r }{(g-r)!} \left< s_\mu,p_3^{g-r} P_{r,n}\right> .
\eeq
The action of $H^{-1}$ on Schur polynomials is given by (see lemma \ref{lemma:H}, property (\ref{H:monomial:Schur}))
\beq
H^{-1}(s_\mu) = \sum_{\substack{|\lambda|=|\mu| \\ \lambda\leq\mu}} \td{K}_{\mu,\lambda} \, m_\lambda .
\eeq
Expressing $P_{r,n}$ in the basis of Schur polynomials (equation \eqref{eqn:hat:Crn}), we get the statement \eqref{eq:main:formula:Agn}. Formula \eqref{eq:main:formula:tau} is just the decomposition in monomial symmetric polynomials of $A_{g,n}$.
\end{proof}

\br
The quantity $Q_{\nu,\mu}$ can also be expressed as a determinant:
\beq\label{eq:Qmunuasdet}
Q_{\nu,\mu}
=
\det_{1\leq i,j\leq n} \left( \frac{\delta_{\hl_{j}(\mu)-\hl_i(\nu)\equiv 0 \pmod{3} \text{ and }\geq 0}}{\bigl(\frac{(\hl_{j}(\mu)-\hl_i(\nu))}{3}\bigr)!}\right) .
\eeq
See appendix \ref{app:Qnumu} for a proof.
\er

\subsection{Proof of the ELO conjecture}
\label{subsec:ELO}

In this section we prove the conjecture of \cite{ELO21}, which consists in remarking that expansion coefficients of the generating polynomials $A_{g,n}$ in the basis of elementary symmetric polynomials manifest some unexpected vanishing.

\subsubsection{Empirical observations}

Let us analyze the homogeneous components of $P_{n}$ (for $n \ge 3$) on the basis of elementary symmetric polynomials $e_\nu$. By separating the powers of $e_1$ from $e_{\nu} = e_{\nu_1}\cdots e_{\nu_{\ell}}$ with $\nu_i\geq 2$, we find a decomposition of the form
\beq\label{eqn:decomp:hatPrn}
P_{r,n} = \sum_{\substack{|\nu|\leq d_{r,n} \\ \nu_i\geq 2}} \mathcal C_{r,n}(\nu) \ e_\nu e_1^{d_{r,n}-|\nu|} .
\eeq

The decomposition of $P_{n}$ in the basis of elementary symmetric polynomials for $n=3$, $4$ and $5$ is displayed in Table \ref{table:hatP:elementary} while that for $n=6$ is displayed in Table \ref{table:hatP6:elementary}. 

\begin{table}[ht]
\centering
\begin{tabular}{r | l}
\toprule 
    $\bm{n=3}$, $\bm{r=0}$ & $P_{0,3}=e_{\emptyset}$ \\[.5ex]
    $\bm{1}$ & $P_{1,3}=\tfrac{1}{2} e_{(3)}$ \\[.5ex]
\midrule
    $\bm{n=4}$, $\bm{r=0}$ & $P_{0,4}=e_{\emptyset}e_1$ \\[.5ex]
    $\bm{1}$ & $P_{1,4}=- \tfrac{3}{2} e_{(4)} + \tfrac{1}{2} e_{(3)} e_1$ \\[.5ex]
    $\bm{2}$ & $P_{2,4}=-\tfrac{1}{4} e_{(4,3)} + \tfrac{1}{12} e_{(4,2)} e_1$ \\[.5ex]
    $\bm{3}$ & $P_{3,4}=\tfrac{1}{24} e_{(4,4,2)}$ \\[.5ex]
\midrule
    $\bm{n=5}$, $\bm{r=0}$ & $P_{0,5} = e_{\emptyset}e_1^2$ \\[.5ex]
    $\bm{1}$ & $P_{1,5}=
    \tfrac{27}{10} e_{(5)} - \tfrac{3}{2} e_{(4)}e_1 + \tfrac{1}{2} e_{(3)}e_1^2$ \\[.5ex]
    $\bm{2}$ & $P_{2,5}=
    \tfrac{6}{5} e_{(5,3)} - \tfrac{3}{10} e_{(5,2)}e_1 + \tfrac{1}{60} e_{(5)}e_1^3 - \tfrac{1}{4} e_{(4,3)}e_1 + \tfrac{1}{12} e_{(4,2)}e_1^2$ \\[.5ex]
    $\bm{3}$ & $P_{3,5}=
    \tfrac{19}{120} e_{(5,5)}e_1 - \tfrac{7}{40} e_{(5,4,2)} + \tfrac{1}{120} e_{(5,4)}e_1^2 + \tfrac{3}{40} e_{(5,3,3)}$ \\[.5ex]
    & $\qquad\quad
    - \tfrac{1}{10} e_{(5,3,2)}e_1 + \tfrac{1}{120} e_{(5,2,2)}e_1^2 + \tfrac{1}{24} e_{(4,4,2)}e_1 $ \\[.5ex]
    $\bm{4}$ & $P_{4,5}=
    \tfrac{1}{36} e_{(5,5,4)} + \tfrac{1}{720} e_{(5,5,3)}e_1 + \tfrac{1}{180} e_{(5,5,2,2)} $ \\[.5ex]
    & $\qquad\quad
    - \tfrac{1}{180} e_{(5,5,2)}e_1^2 - \tfrac{1}{180} e_{(5,4,3,2)} + \tfrac{1}{240} e_{(5,4,2,2)}e_1 $ \\[.5ex]
    $\bm{5}$ & $P_{5,5}=
    - \tfrac{1}{1440} e_{(5,5,5,2)} - \tfrac{1}{480} e_{(5,5,4,2)}e_1 + \tfrac{1}{1440} e_{(5,5,3,2,2)} $ \\[.5ex]
    $\bm{6}$ & $P_{6,5}=
    - \tfrac{1}{720} e_{(5,5,5,5)} + \tfrac{1}{2880} e_{(5,5,5,3,2)} $ \\[.5ex]
\bottomrule
\end{tabular}
\caption{The homogeneous components of $P_n$ for $n = 3, 4, 5$ in the basis of elementary symmetric polynomials.}
\label{table:hatP:elementary}
\end{table}

We observe empirically on these examples the following patterns:
\begin{itemize}
\item 
Observe that some coefficients are repeated for different values of $n$. For instance, $\mathcal C_{0,3}(\emptyset) = \mathcal C_{0,4}(\emptyset) = \mathcal C_{0,5}(\emptyset) = 1$, $\mathcal C_{1,3}((3))=\mathcal C_{1,4}((3))=\mathcal C_{1,5}((3))=\tfrac{1}{2}$, and $\mathcal C_{2,4}((4,2))=\mathcal C_{2,5}((4,2))=\tfrac{1}{12}$, etc. This suggests that the coefficient of a given $e_\nu$ (times the appropriate power of $e_1$) is independent of $n$:
\beq
\mathcal C_{r,n}(\nu) \overset{?}{=} \mathcal C_{r}(\nu) .
\eeq
\item
Notice that $P_{r,5} = e_1 P_{r,4} + \,$remainder and $P_{r,4} = e_1 P_{r,3} + \,$remainder.  This suggests that, more generally:
\beq
P_{r,n} \overset{?}{=} e_1 P_{r,n-1} + \,\text{remainder}.
\eeq

\item
In the decomposition of $P_{r,3}$, $P_{r,4}$ and $P_{r,5}$ only partitions of length $\ell(\nu) \le r$ appear. This suggests the following vanishing property:
\beq
\mathcal C_{r}(\nu) \overset{?}{=} 0
\qquad \text{for} \qquad \ell(\nu) > r.
\eeq
\end{itemize}

We shall prove below that these empirical observations are in fact always true. The first observation follow from a previous result of \cite[proposition 1.2]{ELO21} on the generating polynomial $A_{g,n}$, and is a consequence of the string equation. The second observation is new, and the third one is a restatement of the main conjecture in \cite[conjecture 1.3]{ELO21}.
A final observation which remains an open question is that many coefficients $\mathcal{C}_{r,n}(\nu)$ in \eqref{eqn:decomp:hatPrn} are actually vanishing, see table \ref{table:AllowedvsAppearingCoeff}.

\subsubsection{Decomposition on elementary polynomials}

We start by recalling the first empirical observation for the generating polynomials.

\bt[Decomposition on elementary \cite{ELO21}]

There exist some coefficients $C_g(\nu)$ such that
\beq
A_{g,n} = \frac{1}{{24}^g }\sum_{\substack{|\nu| \leq d_{g,n} \\ \nu_i\geq 2}}
C_{g}(\nu) \ e_\nu e_1^{d_{g,n}-|\nu|}
\eeq
and the coefficients $C_g(\nu)$ are independent of $n$, they depend only on the partition $\nu$.
\et
\begin{proof}
This was proven in \cite{ELO21}.
Let us recall the proof here for completeness.



Just by decomposing on the basis of elementary polynomials, there exist some coefficients $C_{g,n}(\nu)$ such that
\beq
\frac{1}{{24}^g }\sum_{\substack{|\nu| \leq d_{g,n} \\ \nu_i\geq 2}}
C_{g,n}(\nu) \ e_\nu e_1^{d_{g,n}-|\nu|}
\eeq
where we separated the powers of $e_1$ from $e_{\nu} = e_{\nu_1} \cdots e_{\nu_{\ell}}$ with $\nu_i\geq 2$. The string equation (pushforward of $\tau_0$ by the forgetful map $\overline{\mathcal M}_{g,n+1} \to \overline{\mathcal M}_{g,n}$) reads
\beq
A_{g,n+1}(u_1,\dots,u_n,0) = (u_1+\dots+u_n) A_{g,n}(u_1,\dots,u_n).
\eeq
This implies
\begin{align}\label{eq:string.eq.on.Agn}
    \begin{cases} 
        \sum \limits_{\substack{|\nu| \leq d_{g,n} \\ 2 \leq \nu_i}}
        C_{g,n+1}(\nu) \, e_\nu(\uu,0) e_1^{d_{g,n+1}-|\nu|} 
        = e_1\sum\limits_{\substack{|\nu| \leq d_{g,n} \\ 2 \leq \nu_i}}
        C_{g,n}(\nu) \, e_\nu(\uu) e_1^{d_{g,n}-|\nu|} ,
        \vspace{3mm}\\
        \sum\limits_{\substack{|\nu| = d_{g,n+1} \\ 2 \leq \nu_i}}
        C_{g,n+1}(\nu) \, e_\nu(\uu,0) e_1^{d_{g,n+1}-|\nu|} = 0 .
    \end{cases} 
\end{align}
Observe that $e_{\nu_i}(u_1, \ldots, u_n,0) = 0$ for $\nu_i \geq n+1$ and $e_{\nu_i}(u_1, \ldots, u_n,0) = e_{\nu_i}(u_1, \ldots, u_n)$.  Therefore,
identifying each term in \eqref{eq:string.eq.on.Agn} we get
\begin{align}\label{eq:recursive.relation.Cgn}
    \begin{cases}
        C_{g,n+1}(\nu) = C_{g,n}(\nu)\ \quad & |\nu| \leq 3g -3 + n \quad
        \text{and} \quad
        2 \leq \nu_i \leq n \,, i \in [2,\ell(\nu)] ,
        \vspace{3mm}\\
        C_{g,n}(\nu) = 0 & |\nu| = 3g -3 + n \quad \text{and} \quad 
        2 \leq \nu_i < n \,, i \in [2,\ell(\nu)] .
    \end{cases}
\end{align}
\end{proof}

\bc
There exist some coefficients $\mathcal C_{r}(\nu)$ such that
\beq\label{eq1:hatPrnCrnu}
P_{r,n} = \frac{1}{{24}^r} \sum_{\substack{|\nu|\leq d_{r,n} \\ \nu_i\geq 2}}
\mathcal C_{r}(\nu) \ e_\nu e_1^{d_{r,n}-|\nu|}
\eeq
and the coefficients $\mathcal{C}_r(\nu)$ are independent of $n$, they depend only on the partition $\nu$.
\ec
\begin{proof}
We have:
\beq
\begin{split}
12^r \ P_{r,n} 
&=
\sum_{g=0}^r
(-1)^{r-g} \, 24^g \, \frac{p_3^{r-g}}{(r-g)!} H(A_{g,n}) \\
&=
\sum_{g=0}^r \sum_{\substack{|\nu|\leq d_{g,n} \\ \nu_i \geq 2}}
C_g(\nu) \, (-1)^{r-g} \frac{p_3^{r-g}}{(r-g)!} H(e_\nu e_1^{d_{g,n}-|\nu|}) \\
&= \sum_{g=0}^r \sum_{\substack{|\nu|\leq d_{g,n} \\ \nu_i \geq 2}}
C_g(\nu) \, (-1)^{r-g} \frac{p_3^{r-g}}{(r-g)!} H(e_\nu ) e_1^{d_{g,n}-|\nu|}.
\end{split}
\eeq
The action of $H$ on elementary symmetric polynomials is given by $H(e_\lambda) = \sum_{\mu \geq \lambda} H_{\lambda,\mu} e_\mu$ (see lemma \ref{lemma:H}, property (\ref{H:elementary})). The coefficients $H_{\lambda,\mu}$ are independent of $n$. To conclude,
\beq
p_3 = e_1^3-3e_1e_2+3e_3
\eeq
also involves only coefficients independent of $n$.
\end{proof}

\bl\label{lemmaPrnrecurn}
There exist some symmetric polynomials $Q_{r,n}$ of $n$ variables such that
\beq\label{eqPrnrecurn}
P_{r,n} = e_1 P_{r,n-1} + e_n Q_{r,n} .
\eeq
\el
\begin{proof}
In \eqref{eq1:hatPrnCrnu}, the sum over $\nu$ is such that $\nu_i\leq n$ because $e_k=0$ if $k\geq n+1$.
Some partitions $\nu$ in the sum may have some (at least one) rows of length $n$, or no row of length $n$.
We thus separate the sum over $\nu$ into two factors: $P_{r,n} = P^{<}_{r,n} + P^{=}_{r,n}$. The term 
\beq
P^{=}_{r,n}
=
\frac{1}{{24}^r} \sum_{\substack{|\nu|\leq d_{r,n} \\ \nu_1=n, \ \nu_i\geq 2}}
\mathcal C_{r}(\nu) \ e_\nu e_1^{d_{r,n}-|\nu|}
\eeq
has a factor $e_n$ and can be written as $P^{=}_{r,n} = e_n Q_{r,n}$. The other term 
\beq
P^{<}_{r,n} = \frac{1}{{24}^r} \sum_{\substack{|\nu|\leq d_{r,n} \\ \nu_1<n, \ \nu_i\geq 2}}
\mathcal C_{r}(\nu) \ e_\nu e_1^{d_{r,n}-|\nu|}
\eeq
is such that all partitions $\nu$ that appear have $\nu_i\leq n-1$, so they appear also in $P_{r,n-1}$, with the same coefficient $\mathcal C_r(\nu)$. We recognize
\beq
e_1P_{r,n-1} 
= \frac{1}{{24}^r}\sum_{\substack{|\nu|\leq d_{r,n-1} \\ \nu_1<n, \ \nu_i\geq 2}}
\mathcal C_{r}(\nu) \ e_\nu e_1^{d_{r,n}-|\nu|-1} e_1 
= P^{<n}_{r,n} .
\eeq
This concludes the proof that
$P_{r,n} = e_1 P_{r,n-1} + e_n Q_{r,n}$.
\end{proof}

\bp[Bounded rows in the elementary basis]
In the decomposition
\beq\label{eqPrnElambda}
P_{r,n} = \sum_{\substack{|\nu|\leq d_{r,n} \\ \nu_i\geq 2, \  \ell(\nu)\leq r}} \mathcal C_{r}(\nu)\ e_\nu e_1^{d_{r,n}-|\nu|}
\eeq
only partitions of length $\ell(\nu) \le r$ appear.
\ep
\begin{proof}
We shall proceed by recursion on $n$.
The case $n=3$ is easy since $P_3=1+\frac{1}{2} e_3$.
Let us now assume that $n\geq 4$ and the proposition holds for $n-1$. We have
\begin{multline}
P_{n}(\uu)
=
\frac{e_n^{n-\frac32} e^{-\frac{p_3}{12}}}{\Delta(u)}
\Asym \left[ \prod_{i=3}^{n-1}  \left(\frac{d}{du_1}-\frac{d}{du_i}\right) 
\prod_{i=2}^{n-2} \prod_{j=i+2}^{n} \left(\frac{d}{du_i}-\frac{d}{du_j}\right) \times \right. \\
\left. \times
\left(
\frac{e^{\frac{p_3}{12}}}{\sqrt{e_n}} \ \Tr \prod_{i=1}^n \td M(u_i)
\right)
\right].
\end{multline}
Each $\td M(u_i)$ is a polynomial with monomials $u_i^{k_i+1}$ and $k_i \in \{-2,-1,0,1\}$. Let us write
\beq
\Tr \prod_{i=1}^n \td M(u_i) = \sum_{k_1,\dots,k_n} C_{k_1,\dots,k_n} \, u_1^{k_1+1}\cdots u_n^{k_n+1}
\eeq
so that we can decompose the polynomial $P_n$ as 
\beq
P_{n}
= \sum_{k_1,\dots,k_n} C_{k_1,\dots,k_n} \, P^{(k_1,\dots,k_n)}_{n} .
\eeq
Here we have defined the symmetric polynomial:
\begin{multline}
P^{(k_1,\dots,k_n)}_{n}(\uu)
=
\frac{e_n^{n-\frac32} e^{-\frac{p_3}{12}}}{\Delta(u)}
\Asym \left[ \prod_{i=3}^{n-1}  \left(\frac{d}{du_1}-\frac{d}{du_i}\right) 
\prod_{i=2}^{n-2} \prod_{j=i+2}^{n} \left(\frac{d}{du_i}-\frac{d}{du_j}\right) \right. \times \\
\left. \times
\left(
\frac{e^{\frac{p_3}{12}}}{\sqrt{e_n}} \ \prod_{i=1}^n u_i^{k_i+1}
\right)
\right] .
\end{multline}
For each such monomial, let us denote for $j\in\{-2,-1,0,1\}$:
\beq
N_j = \#\{ i  \ : \ k_i=j \}.
\eeq
The terms with $k_i=-1$ come from the top-right corner of the matrix $\td M(u_i)$, the terms with $k_i=-2$ or $k_i=1$ come from the bottom-left corner, and the terms $k_i=0$ from the diagonal. In order for the trace of product of matrices to be non-vanishing, we need that each bottom-left gets paired with a top-right. Therefore we must have
\beq
N_{-1}=N_1+N_{-2} .
\eeq
This implies that
\beq
\sum_{i=1}^n k_i = N_1-N_{-1}-2N_{-2} = -3 N_{-2}.
\eeq
Also, observe that the operators $(d/du_i-d/du_j)$ act as if $e_1$ was a constant. By writing
\beq
p_3 = e_1^3 - 3e_1 e_2 + 3 e_3
\eeq
we see that the term $e^{\frac{e_1^3}{12}}$ passes through the differential operator untouched and cancels out with the corresponding prefactor. We thus obtain
%
%
\begin{multline}\label{eq:polyappearinginPn}
P^{(k_1,\dots,k_n)}_{n}(\uu)
= \frac{e_n^{n-\frac32} e^{\frac{1}{4}e_1 e_2}e^{-\frac{1}{4} e_3}}{\Delta(u)}
\Asym \left[ \prod_{i=3}^{n-1}  \left(\frac{d}{du_1}-\frac{d}{du_i}\right) 
\prod_{i=2}^{n-2} \prod_{j=i+2}^{n} \left(\frac{d}{du_i}-\frac{d}{du_j}\right) \right. \\
\left.
\left(
e^{-\frac{1}{4}e_1 e_2}e^{\frac{1}{4} e_3} \prod_{i=1}^n u_i^{k_i+\frac12}
\right)
\right] .
\end{multline}
The derivatives act either on the exponentials or on the monomials. When they act on the exponentials, they bring down derivatives of $e_2$ or $e_3$.
Let us record:
\begin{itemize}
    \item $b=$ the total number of times a derivative acts on $e^{-\frac14 e_1 e_2}$,
    \item $c=$ the total number of times a derivative acts on $e^{\frac14 e_3}$.
\end{itemize}
Such a term is a homogeneous symmetric polynomial of total degree:
\beq
\begin{split}
\deg 
&= n \left(n-\frac32 \right) - \frac{n(n-1)}{2} + 2b+2c - \left(\frac12 n(n-3)-b-c \right) + \sum_{i=1}^n \left( k_i+\frac12 \right) \\
&= 3b+3c+n + \sum_{i=1}^n k_i \\
&= 3b+3c+n -3 N_{-2} .
\end{split}
\eeq
This homogeneous term contributes to $P^{(k_1,\dots,k_n)}_{r,n}$ if and only if
\beq
3r-3+n = 3b+3c+n-3 N_{-2} .
\eeq
Hence, $r$ satisfies
\beq
r = b+c+1-N_{-2}.
\eeq
Consider the highest possible power of a given $u_i$ in such terms (with fixed $k_1,\dots,k_n$ and fixed $b,c$), with the powers of $e_1$ factored out. It is obtained by acting the least possible with derivatives on monomials. Acting by a $(d/du_j-d/du_m)$ on $e^{\frac14 e_3}$ brings down a polynomial of degree 1 in $u_i$, thus raises the degree by $c$.
Acting by a $(d/du_j-d/du_m)$ on $e^{-\frac14 e_1 e_2}$ brings down a polynomial of the form $e_1$ times a polynomial of degree 1 in $u_i$ only if $j=i$ or $m=i$. Factoring out $e_1$, this may raise the degree at most by $\min(b,n-3)$. We thus have
\beq
\begin{split}
\max\deg_{u_i} 
&\leq \left( n-\frac32 \right)-(n-1)+c +\min(b,n-3) + k_i+\frac12 \\
&\leq c+\min(b,n-3)+k_i .
\end{split}
\eeq
We remark that $e_\nu$ is a symmetric polynomial where each variable can appear at most with power $\ell(\nu)$. Conversely, a symmetric polynomial where powers of $u_i$'s are bounded by $l$ can be written as a linear combination of $e_\nu$ with $\ell(\nu)\leq l$.
This implies that monomials appearing in \eqref{eq:polyappearinginPn} can only contribute to some $e_\nu$ such that $\ell(\nu)\leq c+\min(b,n-3)+k_i$.
Let us compute
\beq
\begin{split}
\ell(\nu)-r 
& \leq c+\min(b,n-3)+k_i - (b+c+1-N_{-2})  \\
& \leq k_i-1 +\min(b,n-3)- b  + N_{-2} \\
& \leq k_i-1 +\min(0,n-3-b) + N_{-2} \\
& \leq k_i-1  + N_{-2} \\
& \leq N_{-2}.
\end{split}
\eeq

If all $k_i\geq -1$ then $N_{-2}=0$. Thanks to this inequality, we conclude that all $e_\nu$ that appear have  $\ell(\nu) \leq r $.

On the other hand, if at least one $k_i=-2$, then $N_{-2} > 0$. However, we can conclude that $\ell(\nu) \leq r $ thanks to the inductive hypothesis.

Indeed, for each monomial in \eqref{eq:polyappearinginPn}, 
the lowest possible power of a given $u_i$ is obtained by acting as much as possible with derivatives on monomials, i.e. at most $(n-3)$ times. In the limit $u_i \rightarrow 0$,
\beq
\min\deg_{u_i} = \left( n-\frac32 \right)-(n-3) + k_i+\frac12 = k_i+2 \geq 0
\eeq
and this bound is reached. Therefore, after we symmetrize on all $u_i$'s, we obtain a symmetric polynomial which does not vanish at $u_i=0$.
This implies that this term is not a factor of $u_1 \cdots u_n = e_n$. Hence, in \eqref{eqPrnrecurn} of lemma \ref{lemmaPrnrecurn}, this term contributes to $e_1 P_{r,n-1}$.
The recursion hypothesis implies that for $P_{r,n-1}$ the $e_\nu$'s that appear all have $\ell(\nu)\leq r$.
\end{proof}

Finally, we can prove the conjecture.

\bt[Conjecture of \cite{ELO21}] \label{thm:Conjecture:ELO21}
In the decomposition
\beq
A_{g,n} = \sum_{\substack{|\nu|\leq d_{g,n} \\ \nu_i\geq 2, \  \ell(\nu)\leq g}} C_{g}(\nu)\ e_\nu e_1^{d_{g,n}-|\nu|}
\eeq
only partitions of length $\ell(\nu) \le g$ appear.
\et
\begin{proof}
We have
\beq
\begin{split}
24^g A_{g,n} 
&=
\sum_{r=0}^{\min(g,\frac{(n-1)(n-2)}{2})} \frac{12^r}{(g-r)!} H^{-1}(p_3^{g-r} P_{r,n}) \\
&=
\sum_{r=0}^{\min(g,\frac{(n-1)(n-2)}{2})}
\sum_{\substack{|\nu|\leq d_{r,n} \\ \nu_i\geq 2, \ \ell(\nu)\leq r}}
\mathcal C_{r}(\nu) \frac{12^r}{(g-r)!}
H^{-1}(p_3^{g-r} e_{\nu} e_1^{d_{r,n}-|\nu|}) \\
&=
\sum_{r=0}^{\min(g,\frac{(n-1)(n-2)}{2})}
\sum_{\substack{|\nu|\leq d_{r,n} \\ \nu_i\geq 2, \ \ell(\nu)\leq r}}
\mathcal C_{r}(\nu) \frac{12^r}{(g-r)!} e_1^{d_{r,n}-|\nu|}
H^{-1}((e_1^3-3e_1 e_2 +3 e_3)^{g-r} e_{\nu} ) \\
&=
\sum_{r=0}^{\min(g,\frac{(n-1)(n-2)}{2})}
\sum_{\substack{|\nu|\leq d_{r,n} \\ \nu_i\geq 2, \ \ell(\nu)\leq r}}
\sum_{a+b+c=g-r}
\mathcal C_{r}(\nu) \frac{12^r (-3)^b 3^c}{a! b! c!}
e_1^{d_{r,n}-|\nu|} H^{-1}(e_1^{3a+b} e_2^b e_3^c  e_{\nu} ) \\
&=
\sum_{r=0}^{\min(g,\frac{(n-1)(n-2)}{2})}
\sum_{\substack{|\nu|\leq d_{r,n} \\ \nu_i\geq 2, \ \ell(\nu)\leq r}}
\sum_{a+b+c=g-r}
\mathcal C_{r}(\nu) \frac{12^r (-3)^b 3^c}{a! b! c!}
e_1^{d_{r,n}-|\nu|+3a+b} H^{-1}(e_2^b e_3^c  e_{\nu} ) .
\end{split}
\eeq
Notice that $e_2^b e_3^c e_{\nu}$ has at most $\ell(\nu)+b+c$ rows of length $\nu_i \ge 2$:
\beq
\ell( e_{\nu,2^b,3^c}) \leq \ell(\nu)+b+c \leq r+(g-r)\leq g.
\eeq
The operator $H^{-1}$ is triangular on the basis of elementary symmetric polynomials (lemma \ref{lemma:H}, property \ref{H:elementary}), which implies that $H^{-1}$ conserves this property. This concludes the proof.
\end{proof}

From the above proof, we also deduce an alternative formula for the generating polynomials.

\bc
The generating polynomial is given by
\begin{multline}
A_{g,n} 
=
\frac{1}{24^g}
\sum_{r=0}^{\min(g,\frac{(n-1)(n-2)}{2})}
\sum_{\substack{|\nu|\leq d_{r,n} \\ \nu_i\geq 2, \ \ell(\nu)\leq r}}
\sum_{a+b+c=g-r} \ \
\sum_{\substack{|\lambda|=|\mu|=|\nu|+3c+2b \\ \lambda \leq\mu \leq (\nu 3^c 2^b)^T}}
\mathcal C_{r}(\nu) \frac{12^r (-3)^b 3^c}{a! b! c!} \times\\
\times
K_{\mu^T,(\nu 3^c 2^b)} \td{K}_{\mu,\lambda}
e_1^{d_{r,n}-|\nu|+3a+b}
m_\lambda .
\end{multline}
\ec
\begin{proof}
The formula follows from the proof of the previous result, and the computation of $H^{-1}(e_{\lambda})$ for $\lambda = (\nu 3^c 2^b)$. This can be easily deduced from the action of $H^{-1}$ on Schur (lemma \ref{lemma:H}, property \ref{H:monomial:Schur}), together with the change of basis from elementary to Schur \eqref{eqn:elementary:to:Schur}:
\beq
H^{-1}(e_{\lambda})
=
\sum_{\substack{|\mu| = |\lambda| \\ \mu \le \lambda^T}} K_{\mu^T,\lambda} H^{-1}(s_\mu)
=
\sum_{\substack{|\nu| = |\mu| = |\lambda| \\ \nu \leq \mu \le \lambda^T}}
K_{\mu^T,\lambda} \td{K}_{\mu,\nu} \, m_{\lambda} .
\eeq
\end{proof}

\subsection{New formulas for the correlators}
\label{subsec:alternative}

In this section, we present some alternative formulations of the main formula in terms of correlators. In particular, we find that the Kostka numbers disappear from the formula computing $W_{g,n}$, by changing basis to that of Schur polynomials. As a consequence, we get an expression for the $n$-point correlators $W_n$ as determinants.

\bt
The correlators are given by
\beq\label{main:formula:corr}
W_{g,n}(\xx)
=
\frac{(-1)^{n} d\xx}{2^{n+1} \, \xx^{3/2}}
\!
\sum_{r=0}^{\min(g,\frac{(n-1)(n-2)}{2})}
\! 12^{r-g} \!
\sum_{\substack{|\nu|=d_{r,n} \\ |\mu|=d_{g,n}}}
\prod_{i=1}^n \frac{\Gamma(\mu_i - i + \frac{5}{2})}{\Gamma(- i + \frac{5}{2})}
\Coeff_{r,n}(\nu) \, Q_{\nu,\mu} \, s_\mu(\xx^{-1})  
\eeq
where we denote $d\xx = dx_1\otimes \dots \otimes dx_n$, $\xx^{3/2} = \prod_{i=1}^n x_i^{3/2}$, and $\xx^{-1} = (x_1^{-1},\dots,x_n^{-1})$.
\et
\begin{proof}
In the definition of the correlators, i.e.
\beq
W_{g,n}(\xx)
\coloneqq (-2)^{-(2g-2+n)}
\sum_{|\lambda| = d_{g,n}}
\braket{  \tau_{\lambda_1}\cdots\tau_{\lambda_n} }_g \prod_{i=1}^n \frac{(2\lambda_i+1)!! dx_i}{2 \, x_i^{\lambda_i+\frac32}} .
\eeq
we substitute the main formula for intersection numbers and, after simplifying the normalization factor $N_{\mu,\lambda}$, we recognize the change of basis from Schur to monomial symmetric:
\beq
\begin{split}
(-1)^n & \frac{2^{n+1} \, \xx^{3/2}}{d\xx} W_{g,n}(\xx) \\
&=
\sum_{r=0}^{\min(g,\frac{(n-1)(n-2)}{2})}
\!\! 12^{r-g} \!\!
\sum_{|\nu|=d_{r,n}} \ \sum_{\substack{|\mu|=|\lambda|=d_{g,n} \\ \mu \ge \lambda}}
\prod_{i=1}^n \frac{\Gamma(\mu_i - i + \frac{5}{2})}{\Gamma(- i + \frac{5}{2})}
\Coeff_{r,n}(\nu) \, Q_{\nu,\mu} \, K_{\mu,\lambda} \, m_\lambda(\xx^{-1}) \\
&=
\sum_{r=0}^{\min(g,\frac{(n-1)(n-2)}{2})}
\!\! 12^{r-g} \!\!
\sum_{\substack{|\nu|=d_{r,n} \\ |\mu|=d_{g,n}}}
\prod_{i=1}^n \frac{\Gamma(\mu_i - i + \frac{5}{2})}{\Gamma(- i + \frac{5}{2})}
\Coeff_{r,n}(\nu) \, Q_{\nu,\mu} \, s_\mu(\xx^{-1}) .
\end{split}
\eeq
\end{proof}


Moreover, using the expression of $Q_{\nu,\mu}$ and $s_{\mu}$ in terms of determinants (see lemma \ref{lem:multPower} and \eqref{eqn:Schur:complete:homog} respectively), we can get a new expression for the $n$-point correlators. 

\bt
The $n$-point correlators are given by
\beq\label{eqn:Wn:det}
W_n(\xx) 
= 
(-1)^{n}
\frac{d\xx}{2^{n+1} \, \xx^{3/2}}
\sum_{r=0}^{\frac{(n-1)(n-2)}{2}} \sum_{|\nu|=d_{r,n}}
\frac{\Coeff_{r,n}(\nu)}{\prod_{i=1}^n \Gamma\left(-i + \frac{5}{2}\right)}
\det \left( F(\nu,\xx^{-1}) \right),
\eeq
where $F = (F_{i,j})$ is the $n \times n$ matrix given by
\beq
F_{i,j}(\nu,\xx)
=
\sum^{\infty}_{k=0} \frac{\Gamma\left(\hl_{i}(\nu) - n + 3k + \frac{5}{2} \right)}{k!12^k} \,
h_{\hl_{i}(\nu) - (n-j) + 3k}(\xx) .
\eeq
\et
\begin{proof}
Starting from \eqref{main:formula:corr} summed over $g$ and exchanging the summations over $g$ and $r$ using $\sum_{g \ge 0} \sum_{r=0}^{\min(g,\frac{(n-1)(n-2)}{2})} = \sum_{r=0}^{\frac{(n-1)(n-2)}{2}} \sum_{g \ge r}$, we find:
\beq
\begin{split}
(-1)^n & \frac{2^{n+1} \, \xx^{3/2}}{d\xx} W_{n}(\xx) \\
&=
\sum_{r=0}^{\frac{(n-1)(n-2)}{2}} \sum_{g \ge r}
12^{r-g}
\sum_{\substack{|\nu|=d_{r,n} \\ |\mu|=d_{g,n}}}
\prod_{i=1}^n \frac{\Gamma(\mu_i - i + \frac{5}{2})}{\Gamma(- i + \frac{5}{2})}
\Coeff_{r,n}(\nu) \, Q_{\nu,\mu} \, s_\mu(\xx^{-1}).
\end{split}
\eeq
We can now express $Q_{\nu,\mu}$ and $s_{\mu}$ as determinants, with the dependence on the partitions $\mu$ and $\nu$ appearing through the quantities $\hl_i(\mu)$ and $\hl_i(\nu)$ only. The same dependence appears in $\Gamma(\mu_i - i + \frac{5}{2}) = \Gamma(\hl_i(\mu)-n+\frac52 )$ and in $(r - g) = \frac{1}{3} \sum_i ( \hl_i(\nu)-\hl_i(\mu) )$. Thus:
\beq\label{eqn:Wn:det:temp}
\begin{split}
(-1)^n & \frac{2^{n+1} \, \xx^{3/2}}{d\xx} W_{n}(\xx) \\
&=
\sum_{r=0}^{\frac{(n-1)(n-2)}{2}}
\frac{\Coeff_{r,n}(\nu)}{\prod_{i=1}^n \Gamma\left(-i + \frac{5}{2}\right)}
\sum_{|\nu|=d_{r,n}} \
\sum_{\substack{|\mu| \ge |\nu| \\ |\nu|-|\mu| \equiv 0 \pmod{3}}} 12^{\frac{1}{2} \sum_{i=0}^n (\hl_i(\nu)-\hl_i(\mu))} \times 
\\
& \qquad \times 
\det
\left(
\frac{\mathfrak{d}_{\hl_j(\mu)-\hl_i(\nu)}}{\bigl( \frac{\hl_j(\mu)-\hl_i(\nu)}{3}\bigr)!}
\right) \prod_{i=1}^n \Gamma\left(\hl_i(\mu)-n +\frac{5}{2}\right) \det \bigl( h_{\hl_i(\mu) - (n-j)}(\xx^{-1}) \bigr).
\end{split}
\eeq
where we set
\beq
\mathfrak{d}_{k}
\coloneqq
\begin{cases}\label{def:mathfrak:d}
    1 & \text{if $k \equiv 0 \pmod{3}$ and $k \ge 0$}, \\
    0 & \text{otherwise}.
\end{cases}
\eeq

As $\mu$ appears only through the quantities $\hl_i(\mu)$ and the conditions $|\mu| \ge |\nu|$ and $|\nu|-|\mu| \equiv 0 \pmod{3}$ are automatically imposed by the determinant (since it is proportional to $\braket{p_3^k s_{\nu},s_{\mu}}$), we can replace $\sum_{\mu}$ with $\sum_{\hl_1 > \cdots  > \hl_n \ge 0}$. Additionally, the above formula is symmetric under $\hl_{i_1}\leftrightarrow \hl_{i_2}$ for any $i_1, i_2 \in [1,n]$ since both determinants will contribute the same sign. Thus, we may replace $\sum_{\hl_1 > \cdots > \hl_n \ge 0}$ with $\frac{1}{n!} \sum_{\hl_1,\ldots,\hl_n \ge 0}$.
Expressing each determinant via the Leibniz formula, we get
\beq
\begin{split}
&\sum_{\mu} 12^{-\frac{1}{3} \sum_{i=0}^n \hl_i(\mu)} 
\det
\left(
\frac{\mathfrak{d}_{\hl_j(\mu)-\hl_i(\nu)}}{\bigl( \frac{\hl_j(\mu)-\hl_i(\nu)}{3}\bigr)!}
\right)  
\prod_{i=1}^n \Gamma\left(\hl_i(\mu)-n +\frac{5}{2}\right)
\det \bigl( h_{\hl_i(\mu) - (n-j)}(\xx^{-1}) \bigr)
\\
&=
\frac{1}{n!} \sum_{\hl_1,\ldots, \hl_n \ge 0} \
\sum_{\sigma, \rho \in \mathfrak{S}_n} 
(-1)^{\sigma \rho} 
\prod_{i=1}^n 12^{-\frac{\hl_i}{3} } \,
\frac{\Gamma \left(\hl_i - n + \frac{5}{2} \right)}{\bigl(\frac{\hl_i - \hl_{\rho(i)}(\nu)}{3}\bigr)!} \,
h_{\hl_i - (n-\sigma(i))}(\xx^{-1}) \, \mathfrak{d}_{\hl_i - \hl_{\rho(i)}(\nu)}
\\
&=
\frac{1}{n!} \sum_{\sigma, \rho \in \mathfrak{S}_n} 
(-1)^{\sigma \rho} 
\prod_{i=1}^n
12^{-\frac{1}{3}\hl_{\rho(i)}(\nu)}
\sum_{k \ge 0} \frac{\Gamma \left(\hl_{\rho(i)}(\nu) - n + 3k + \frac{5}{2}\right) }{12^k k!} \,
h_{\hl_{\rho(i)}+3k - (n-\sigma(i))}(\xx^{-1})
\\
&=
\prod_{i=1}^n 12^{-\frac{1}{3} \hl_i(\nu)} \, \det\bigl( F(\nu,\xx^{-1}) \bigr) .
\end{split}
\eeq
In the last equality, we performed the sum over $\rho$ by relabelling $\rho(i) \to i$. Moreover, the prefactor $\prod_{i=1}^n 12^{-\frac{1}{3} \hl_i(\nu)}$ cancels out when inserted in \eqref{eqn:Wn:det:temp}, and we thus obtain \eqref{eqn:Wn:det}.
\end{proof}

\section{Algorithmic complexity}
\label{sec:cmlxty}

We are interested in the computational complexity of \eqref{eqn:int:numbers} with respect to the genus $g$ at fixed $n$, for $g$ large. We consider that an addition or multiplication has cost 1, so that for example the computational complexity of $n!$ is $n$.

Before proceeding with the estimate of the algorithmic complexity of the main formula, let us have a closer look at the computation of Kostka numbers. It is known that the complexity of computing Kostka numbers is a $\#$P-complete hard problem \cite{Led05, Nar06}, as a function of the weight. However, here we are interested in Kostka numbers with partitions having a bound $n$ on the number of rows, and then the problem is polynomial \cite{Led05, Nar06}. Indeed, from the following formula for the Kostka numbers:
\beq
K_{\mu,\lambda} = \sum_{\sigma\in \mathfrak S_n} (-1)^\sigma N_\mu(\lambda+(\sigma)-(n))
\eeq
where $(n) = (1,\dots,n)$, $(\sigma) = (\sigma(1),\dots,\sigma(n))$, and
\beq    
N_{\mu}(\nu) = \#\left\{ M \in \textrm{Mat}_{n\times n}(\mathbf N) \ : \ \sum_j M_{i,j} = \mu_i \ , \ \sum_i M_{i,j} = \nu_j  \right\},
\eeq
we see that
\beq
\begin{split}
\text{complexity of }K_{\mu,\lambda}
&\leq n! \ \prod_{i=1}^n \binom{n+\mu_i}{n} \\
&\leq n! \ \prod_{i=1}^{\ell(\mu)} \frac{\mu_i^n}{n!} \ e^{\frac{n(n+1)}{2\mu_i}} \\
&\leq (n!)^{1-n} e^{\frac{n^2(n+1)}{2}} \ \Biggl( \prod_{i=1}^{\ell(\mu)} \mu_i \Biggr)^n \\
&\leq (n!)^{1-n} e^{\frac{n^2(n+1)}{2}} n^{-n^2} \ |\mu|^{n^2}.
\end{split}
\eeq
For $|\mu|=3g-3+n$, this number grows as $O(g^{n^2})$. This implies the following result.

\bp
The computational complexity of the intersection number $\braket{\tau_\lambda}_g$ from formula \eqref{eqn:int:numbers} is at most $O(g^{n^2+n})$ at large $g$, independently of $\lambda$. Moreover, the most expensive computation is that of Kostka numbers.
\ep
\begin{proof}
Let us analyze the computational complexity of each term separately.
\begin{itemize}[wide]
    \item The coefficients $24^g$ have a complexity of order $O(g)$.

    \item We have a sum over $r$, whose number of terms is quadratic in $n$ and independent of $g$ (for large $g$). The sum over $\nu$ is also independent of $g$. Therefore, the sum over $r$ and $\nu$ can be neglected in the analysis. Similarly, the polynomial $P_{r,n}$ is independent of $g$, and so are the coefficients $\Coeff_{r,n}(\nu)$.

    \item The number of partitions of weight $d$, with at most $n$ rows is estimates as:
    \beq
    \#\{\mu \ : \ |\mu|=d, \ \ell(\mu)\leq n \}
    \leq
    \frac{(d+n)!}{n! d!} \leq \frac{d^n}{n!} e^{\frac{n(n+1)}{2d}}
    \sim
    \frac{d^n}{n!} \quad \text{ at large }d .
    \eeq
    As $d= 3g-3+n$, this number grows like $O(g^n)$. This proves that the sum over $\mu$ with weight $|\mu|=3g-3+n$ and at most $n$ rows is of order $O(g^{n})$.

    \item The coefficients $Q_{\nu,\mu}$ are determinants of $n \times n$ matrices (cf. appendix \ref{app:Qnumu}), i.e. sum of $n!$ terms. Each term is a product of factorials, and the total number of factors is $(g-r)$. The algorithmic complexity is at most $n! (g-r)$, i.e. the coefficients $Q_{\mu,\nu}$ has complexity of order $O(g)$.

    \item The Kostka numbers $K_{\mu,\lambda}$ with bounded rows have a complexity of order $O(g^{n^2})$.

    \item The normalization coefficient $N_{\mu,\lambda}$ depends on both $\mu$ and $\lambda$ partitions of $3g-3+n$. The $\lambda$-dependence enters as a product of terms whose number is bounded by $3g-3+n$. The $\mu$-dependence enters as a product of factorials. Each factorial is a product of factors given by the row length $|\mu_i-i|$. The total number of products is bounded by $3g-3+n+n(n+1)/2$. All together, the coefficient $N_{\mu,\lambda}$ has complexity of order $O(g)$.


\end{itemize}
Therefore the total complexity is of order
\beq
O(g) + O(g^n) \left(O(g) + O(g^{n^2}) + O(g) \right) = O(g^{n^2+n}) .
\eeq
\end{proof}

Let us mention that these bounds are very large, probably much overestimated, and in practice the number of operations is much smaller.

\section{Conclusion}
\label{sec:conclusion}


We proposed a formula for intersection numbers that involves only sums over partitions of combinatorial factors.  It involves no algebro-geometric integral, no solving of KdV equations, no recursion, no differential equations. Additionally, let us emphasize that the application of our formula leads to the proof of theorem \ref{thm:Conjecture:ELO21}, i.e. a proof of the conjecture of \cite{ELO21}.

Moreover, the number of terms in the sums is independent of the genus $g$. The $g$-dependence is entirely coming from the $e^{\frac{p_3}{12}}$ and is easily taken into account by the formula.
Altogether, this means that there is an underlying structure of intersection numbers, showing that they are encoded by a far smaller set of coefficients than it would seem at first sight.

Theorem \ref{thm:Conjecture:ELO21}, implies that all partitions with length $\ell(\mu)> g$ have vanishing coefficients in the decomposition in the basis of elementary symmetric polynomials.
Actually, we also observe empirically, that there are even more partitions that have vanishing coefficients than the length restriction implies.
For example, consider the expansion of $P_6$ in terms of elementary symmetric polynomials: the homogeneous component $P_{10,6}$ involves only $4$ partitions that appear with non-zero coefficient, while the allowed ones (by length $\leq r$ and degree $\leq d_{r,n}$) are $187$.
This amazing suppression of many expected terms evokes a deep hidden structure of intersection numbers that should be further investigated.


Finally, we also expect that this closed formula could possibly be employed to compute large $g$ or large $n$ asymptotics, and can be implemented in practical computational algorithms.

\section*{Acknowledgements}

This work was funded by the ERC Synergy grant \textbf{ReNewQuantum ERC-2018-SyG 810573}.
D.~M. acknowledges support by Onassis Foundation under scholarship ID: F ZR 038/1-2021/2022.
We also thank E. Garcia-Failde, A. Giacchetto, M. Kontsevich, D. Lewański for discussions. 

\bigskip\bigskip

\hrule
\appendix{}
\makeatletter
\renewcommand{\@seccntformat}[1]{Appendix \csname the#1\endcsname\quad}
\makeatother
\renewcommand{\thesection}{\Alph{section}}
\numberwithin{equation}{section}

\section{Trace and identity}
\label{ApplemmaTraceId}

\bl
Let $M_1,\dots, M_n$ be some matrices. Then
\beq\label{eqn:shift:Id}
\sum_{\sigma\in \mathfrak S_n^{\textup{cycl}}} \frac{\Tr \prod_{i=1}^n M_{\sigma^i(1)}}{\prod_{i=1}^n (x_i-x_{\sigma(i)})}
\eeq
is invariant under $M_i\to M_i+\alpha_i \Id$ for any $\alpha_i \in \mathbb{C}$.
\el

\begin{proof}
Without loss of generality, we can assume that $\alpha_2 = \cdots = \alpha_n = 0$. Equation \eqref{eqn:shift:Id} defines a polynomial of degree $1$ in $\alpha_1$, with leading coefficient being a rational function of $x_1$ with simple poles at $x_1 = x_i$ for $i = 2,\dots,n$.
Since it is symmetric in the variables $x_2,\dots,x_n$, let us compute the residue at $x_1 = x_2$ (the residues at $x_1 = x_i$ for $i = 3,\dots,n$ have the same value). The only terms that can contribute to it are those for which $\sigma(1) = 2$ and $\sigma^{-1}(1) = 2$. Let $\mathfrak{S}_+$ the subset of permutations $\sigma \in \mathfrak{S}_n^{\textup{cycl}}$ such that $\sigma(1) = 2$ and $\mathfrak{S}_-$ the set of permutations $\sigma$ such that $\sigma(2)=1$.

For any permutation $(1 \ 2 \ i_3 \ldots i_n) \in \mathfrak{S}_{+}$ there exists a permutation $(2 \ 1 \ i_3 \ldots i_n) \in \mathfrak{S}_{-}$. In other words $\forall \sigma \in \mathfrak{S}_{+}$, $\exists \tau \in \mathfrak{S}_{-} : \sigma = \rho \tau \rho \,, \text{where} \, \rho=(1,2).$ Therefore,
\begin{align}
A 
&\coloneqq
\sum_{\sigma\in \mathfrak{S}_+}  \frac{\Tr M_2 \prod_{i=2}^{n-1} M_{\sigma^i(1)}}{\prod_{i=1}^n (x_i-x_{\sigma(i)})} \nonumber
\\
&=
\sum_{\sigma\in \mathfrak{S}_-}  \frac{\Tr \Big( M_2 \prod_{i=2}^{n-1} M_{(\rho \tau \rho)^i(1)} \Big) }{\prod_{i=1}^n (x_i-x_{(\rho \tau \rho)(i)})} 
=
\sum_{\sigma\in \mathfrak{S}_-}  \frac{\Tr \Big( M_2 \prod_{i=1}^{n-2} M_{\tau^i(1)} \Big) }{\prod_{i=1}^n (x_i-x_{(\rho \tau \rho)(i)})} 
\nonumber
\\
&=
\sum_{2<i_3< \cdots <i_n \leq n} \frac{\Tr \Big( M_2 \prod_{j=3}^{n} M_{i_j} \Big) }{(x_1 - x_2)(x_2 - x_{i_3})(x_{i_n} - x_1)\prod_{j=3}^n (x_{i_j} - x_{i_j +1})} ,
\end{align}
\begin{align}
B & \coloneqq
\sum_{\tau \in \mathfrak{S}_-}   \frac{\Tr \Big( M_2 \prod_{i=1}^{n-2} M_{\tau^i(1)} \Big)}{\prod_{i=1}^{n} (x_i-x_{\tau(i)})} 
\nonumber \\
&=
\sum_{2<i_3< \cdots <i_n \leq n}
\frac{\Tr \Big(M_2 \prod_{j=3}^{n} M_{i_j}\Big)}{(x_1 - x_{i_3})(x_2 - x_1)(x_{i_n} - x_2)\prod_{j=3}^{n} (x_{i_j} - x_{i_j +1})} .
\end{align}
Adding $A$ and $B$, we find
\begin{align}
A+B 
&= 
\sum_{2<i_3< \cdots <i_n \leq n} \frac{x_{i_3}-x_{i_n}}{(x_2 - x_{i_3})(x_{i_n} - x_1)(x_1 - x_{i_3})(x_{i_n} - x_2)}
\frac{\Tr \Big(M_2 \prod_{j=3}^{n} M_{i_j}\Big)}{\prod_{j=3}^{n} (x_{i_j} - x_{i_j +1})} .
\end{align}
Notice that no pole of the form $(x_1-x_2)$ appears. Thus, 
\begin{align}
\Res_{x_1 \rightarrow x_2} \big( A+B \big) = 0.
\end{align} 
Additionally, 
\begin{align}
\lim_{x_1 \rightarrow \infty} \big( A+B \big) = 0.
\end{align}
Thus the coefficient of $\alpha_1$ is zero. This concludes the proof.
\end{proof}

\section{Properties of the operator \texorpdfstring{$H$}{H}}
\label{app:H}

In this appendix we shall prove the main properties of the operator $H$ (lemma \ref{lemma:H}). We recall here its definition:
\beq
H(f(\uu)) = \frac{e_n^{n-\frac32}}{\Delta(\uu) D_n} \ \Delta(d/d\uu) \left( \sqrt{e_n} \ f(\uu) \right)
\eeq
with
\beq
\Delta(d/d\uu) \coloneqq \prod_{i<j} \left(\frac{d}{du_i} - \frac{d}{du_j} \right)
\quad,\quad
D_n \coloneqq \frac{G(n-\frac12)}{G(\frac12) \Gamma(-\frac12)^{n-1}} .
\eeq

\paragraph{Property (\ref{H:end:deg}).} The operator $H$ is clearly linear and it preserves the degree. We will deduce that $H$ is invertible on the space of symmetric polynomials from the first part of property (\ref{H:monomial:Schur}).

\paragraph{Property (\ref{H:e1}).} The fact $H(1) = 1$ is a specialization of property (\ref{H:monomial:Schur}). The property $H(e_1^k f) = e_1^k H(f)$ follows from the fact that multiplication by $(\frac{d}{du_i} - \frac{d}{du_j}) e_1^k = 0$.

\paragraph{Property (\ref{H:monomial:Schur}).} We first compute $H(m_{\lambda})$. From the definition of the monomial symmetric polynomials and the expression $\Delta(d/d\uu) = \det (d/du_i)^{n-j} = \sum_{\sigma} (-1)^\sigma \prod_i (d/du_i)^{n-\sigma(i)}$, we deduce:
\beq
\begin{split}
D_n z_\lambda H(m_\lambda) 
& = \frac{1}{\Delta(\uu)}\Asym \left[
    \sum_{\sigma \in \mathfrak{S}_n} (-1)^\sigma
    \prod_{i=1}^n u_i^{n-\frac32} \left( \frac{d}{du_i} \right)^{n-\sigma(i)} u_i^{\lambda_i+\frac12}
    \right] \\
& = \frac{1}{\Delta(\uu)}\Asym \left[
    \sum_{\sigma \in \mathfrak{S}_n} (-1)^\sigma
    \prod_{i=1}^n \frac{\Gamma(\lambda_i+\frac32)}{\Gamma(\lambda_i+\frac32 - n+\sigma(i))} \  u_i^{\lambda_i+\sigma(i)-1}
    \right] .
\end{split}
\eeq
The antisymmetrization kills all the terms such that  all $\lambda_i+\sigma(i)-1$ are not distinct. In this case, we order them by a permutation $\rho$ such that
\beq
 \hl_{\rho(i)} = \lambda_i+\sigma(i)-1 
\quad , \quad
\hl_1>\hl_2>\dots>\hl_n\geq 0.
\eeq
Set $\mu_i \coloneqq \hl_i+i-n$, which satisfies $|\mu| = |\lambda|$. We also recognize the Schur polynomial $ \frac{1}{\Delta(\uu)}\Asym({\prod_{i=1}^n u_i^{\hl_{\rho(i)}} }) = (-1)^\rho s_\mu  $:
\beq
D_n z_\lambda H(m_\lambda) 
=
\sum_{|\mu| = |\lambda|} \sum_{\rho,\sigma \in \mathfrak{S}_n}
(-1)^{\sigma\rho}
\prod_{i=1}^n \frac{\Gamma(\lambda_i+\frac32)}{\Gamma(\mu_i-i+\frac52)}
\delta_{\lambda_i+\sigma(i)-1,\mu_{\rho(i)}-\rho(i)+n} \ s_\mu.
\eeq
Upon changing $\sigma(i) = n+1-\td\sigma(i)$, where $(-1)^{\td\sigma} = (-1)^\sigma (-1)^{\frac{n(n-1)}{2}}$, we get
\beq
D_n z_\lambda H(m_\lambda) 
= (-1)^{\frac{n(n-1)}{2}}
\sum_{|\mu| = |\lambda|} \sum_{\rho,\td\sigma \in \mathfrak{S}_n}
(-1)^{\td\sigma\rho}
\prod_{i=1}^n \frac{\Gamma(\lambda_i+\frac32)}{\Gamma(\mu_i-i+\frac52)}
\delta_{\lambda_i+n-\td\sigma(i),\mu_{\rho(i)}-\rho(i)+n} \ s_\mu .
\eeq
We recognize the normalized inverse matrix of Kostka numbers $\td{K}^{-1} = (\td{S}_{\lambda,\mu})$, see \eqref{eqn:S:det}, and thus
\beq
H(m_\lambda) 
= (-1)^{\frac{n(n-1)}{2}} \frac{1}{D_n} 
\prod_{i=1}^n \frac{\Gamma(\frac32)}{\Gamma(-i+\frac52)}
\sum_{\substack{|\mu|=|\lambda| \\ \mu\leq\lambda}}
    \td{S}_{\lambda,\mu} \, s_\mu .
\eeq
Let $r_n \coloneqq (-1)^{\frac{n(n-1)}{2}} D_n \prod_{i=1}^n \frac{\Gamma(-i+\frac52)}{\Gamma(\frac32)}$. We have $r_1 = D_1 = 1$, and
\beq
\begin{split}
\frac{r_{n+1}}{r_n} 
&= (-1)^{n} \frac{G(n+\frac12)}{G(n-\frac12) \Gamma(-\frac12)} \frac{\Gamma(-n+\frac32)}{\Gamma(\frac32)} \\
&= (-1)^{n} \frac{\Gamma(n-\frac12)\Gamma(-n+\frac32)}{\Gamma(-\frac12)\Gamma(\frac32)} \\
&= (-1)^{n} \frac{\Gamma(n-\frac12)\Gamma(1-(n-\frac12))}{(-2)\Gamma(\frac12)\frac12\Gamma(\frac12)} \\
&= (-1)^{n-1} \frac{\pi}{\pi \sin((n-\frac12)\pi) } \\
&= 1,
\end{split}
\eeq
which implies $r_n = 1$ for all $n$. This proves the wanted formula:
\beq\label{eqn:H:monomial}
H(m_\lambda) = \sum_{\substack{|\mu|=|\lambda| \\ \mu\leq\lambda}} \td{S}_{\lambda,\mu} \, s_\mu .
\eeq
From the above formula we deduce that $H$ is an endomorphism: it send the basis of monomial symmetric polynomials to that of Schur, and the matrix $(\td{S}_{\lambda,\mu})$ realizing this is invertible. Moreover, we can deduce that $H(1) = 1$. The formula for $H^{-1}(s_{\mu})$ then easily follows:
\beq
H^{-1}(s_\mu) = \sum_{\substack{|\lambda|=|\mu| \\ \lambda\leq\mu}} \td{K}_{\mu,\lambda} \, m_\lambda .
\eeq

\paragraph{Property (\ref{H:elementary}).} In order to compute $H(e_{\lambda})$, we can express $e_{\lambda}$ in the basis of monomial symmetric polynomials (see \eqref{eqn:elementary:to:Schur} and \eqref{eqn:Kostka}), compute $H$ on monomial symmetric using \eqref{eqn:H:monomial}, and return to elementary:
\beq
\begin{split}
  H(e_\lambda)
  &=
  \sum_{\tau\leq \lambda^T} K_{\tau^T,\lambda} H(s_{\tau}) \\
  &=
  \sum_{\sigma\leq \tau \leq \lambda^T} K_{\tau^T,\lambda} K_{\tau,\sigma} H(m_{\sigma}) \\
  &=
  \sum_{\rho \leq \sigma \leq \tau \leq \lambda^T} K_{\tau^T,\lambda} K_{\tau,\sigma} \td{S}_{\sigma,\rho} \, s_\rho \\
  &=
  \sum_{\mu^T \leq \rho \leq \sigma \leq \tau \leq \lambda^T} K_{\tau^T,\lambda} K_{\tau,\sigma} \td{S}_{\sigma,\rho} S_{\mu,\rho^T} \, e_{\mu} .
\end{split}
\eeq
The coefficient of $e_\mu$ is non-zero only if $\mu^T\leq\rho\leq\sigma\leq\tau\leq \lambda^T$, thus $\mu^T\leq \lambda^T$, i.e. if $\mu \geq \lambda$. The matrix $(H_{\lambda,\mu})$ is thus upper triangular. Moreover, the Kostka numbers, as well as their inverse and the normalization coefficients $N_{\rho,\sigma}$, do not depend on $n$. Hence, the same holds for $H_{\lambda,\mu}$.

To conclude, consider the case $\lambda=(k,1^l)$. Since $H(e_k e_1^l) = e_1^l H(e_k)$, it suffices to prove \eqref{eqn:H:hook} in the case where $\lambda$ is a single line of length $\lambda=(k)$.
This implies that $\lambda^T$ is minimal, thus we must have $\mu^T=\rho=\sigma=\tau=\lambda^T =(1)^k$. Since $K_{\nu,\nu}=S_{\nu,\nu}=1$, we get
\beq
H(e_k) = \frac{1}{N_{(1)^k,(1)^k}} e_k = (-1)^k \frac{3^{k-1}}{(2k-5)!!} e_k.
\eeq

\section{An alternative expression for \texorpdfstring{$Q_{\nu,\mu}$}{Q}}
\label{app:Qnumu}

\bl\label{lem:multPower}
Let $\mu$, $\nu$ partitions of length $\leq n$ such that $|\mu|-|\nu|=3k$ is a positive multiple of $3$.
Then we have
\beq\label{eqQnumu}
Q_{\nu,\mu}
\coloneqq
\frac{1}{k!}
\left< p_3^{k} s_\nu,s_\mu \right>
    = \det_{1\leq i,j\leq n} \left( \frac{\mathfrak{d}_{\hl_{j}(\mu)-\hl_i(\nu)}}{\bigl( \frac{\hl_{j}(\mu)-\hl_i(\nu)}{3} \bigr)!}\right) ,
\eeq
where $\mathfrak{d}_k$ is defined in \eqref{def:mathfrak:d}. It is equal to 1 if $k \equiv 0 \pmod{3}$ and $k \ge 0$, and $0$ otherwise.
\el
\begin{proof}
From the definition of the scalar product, we have
\beq
\begin{split}
\frac{1}{k!}
\braket{ p_3^{k} s_\nu,s_\mu}
&=
\frac{1}{k! \, n!} \prod_{i=1}^n \Res_{u_i\to 0} \frac{du_i}{u_i} 
p_3^k \det u_i^{\hl_{j}(\nu)} \det u_i^{-\hl_{j}(\mu)} \\
&=
\frac{1}{k! \, n!}
\sum_{d_1+\dots+d_n=k}
\frac{k!}{\prod_{i=1}^n d_i!} \prod_{i=1}^n \Res_{u_i\to 0}
\frac{du_i}{u_i} u_i^{3d_i} \det u_i^{\hl_{j}(\nu)} \det u_i^{-\hl_{j}(\mu)} \\
&=
\frac{1}{n!}
\sum_{d_1+\dots+d_n=k} \ \sum_{\sigma,\rho \in \mathfrak S_n}
(-1)^{\sigma\rho} \prod_{i=1}^n \frac{1}{d_i!} \Res_{u_i\to 0} \frac{du_i}{u_i} 
u_i^{3d_i} u_i^{\hl_{\rho(i)}(\nu)} u_i^{-\hl_{\sigma(i)}(\mu)} \\
&=
\frac{1}{n!}
\sum_{d_1+\dots+d_n=k} \ \sum_{\sigma,\rho \in \mathfrak S_n}
(-1)^{\sigma\rho} \prod_{i=1}^n \frac{1}{d_i!} \delta_{3d_i+\hl_{\rho(i)}(\nu),-\hl_{\sigma(i)}(\mu)}.
\end{split}
\eeq
The sum over $d_i$ actually reduces to at most one term because of the Kronecker deltas, and we must have
\beq
d_i = \frac{\hl_{\sigma(i)}(\mu)-\hl_{\rho(i)}(\nu)}{3},
\eeq
which is possible if and only if the RHS is a positive integer. The fact that $\sum_i d_i=k$ is then automatically satisfied since $\sum_i \hl_i(\mu)-\hl_i(\nu)=|\mu|-|\nu|=3k$. This gives
\beq
\frac{1}{k!}
\left< p_3^{k} s_\nu,s_\mu \right>
=
\frac{1}{n!} \sum_{\sigma,\rho \in \mathfrak S_n}
(-1)^{\sigma\rho} \prod_{i=1}^n \frac{\mathfrak{d}_{\hl_{\sigma(i)}(\mu)-\hl_{\rho(i)}(\nu)}}{\bigl( \frac{\hl_{\sigma(i)}(\mu)-\hl_{\rho(i)}(\nu)}{3} \bigr)!}  .
\eeq
By relabeling $i \to \rho(i)$, we get rid of $\rho$-sum and obtain
\beq
\begin{split}
\frac{1}{k!}
\left< p_3^{k} s_\nu,s_\mu \right>
&=
\sum_{\sigma \in \mathfrak S_n} (-1)^{\sigma}
\prod_{i=1}^n \frac{\mathfrak{d}_{\hl_{\sigma(i)}(\mu)-\hl_{i}(\nu)}}{\bigl( \frac{\hl_{\sigma(i)}(\mu)-\hl_{i}(\nu)}{3} \bigr)!} \\
&= \det \left( \frac{\mathfrak{d}_{\hl_{j}(\mu)-\hl_{i}(\nu)}}{\bigl( \frac{\hl_{j}(\mu)-\hl_{\rho(i)}(\nu)}{3} \bigr)!} \right).
\end{split}
\eeq
\end{proof}

\section{\texorpdfstring{$P_6$}{P6} in the basis of \texorpdfstring{$e_{\mu}$'s}{e's} and number of non-zero coefficients}
\label{app:hatP6}

\begin{center}
\begin{longtable}{r|l}
	\toprule
	$\bm{r=0}$ & $P_{0,6} = e_{\emptyset} e_1^{3}$  \\[.5ex]
    $\bm{1}$ & $P_{1,6}= 
    \tfrac{1}{2} e_{(3)} e_1^{3} 
    - \tfrac{3}{2} e_{(4)} e_1^{2} 
    + \tfrac{27}{10} e_{(5)} e_1 
    - \tfrac{162}{35} e_{(6)}
    $ \\[.5ex]
    $\bm{2}$ & $P_{2,6}=
    \tfrac{1}{12} e_{(4,2)} e_1^{3} 
    + \tfrac{1}{60} e_{(5)} e_1^{4} 
    - \tfrac{9}{140} e_{(6)} e_1^{3} 
    - \tfrac{1}{4} e_{(4,3)} e_1^{2} 
    - \tfrac{3}{10} e_{(5,2)} e_1^{2}
    $ \\[.5ex]     
	&
    $\qquad \quad     
    + \tfrac{6}{5} e_{(5,3)} e_1 
    + \tfrac{93}{140} e_{(6,2)} e_1 
    - \tfrac{513}{140} e_{(6,3)}
    $ \\[.5ex]
    $\bm{3}$ & $P_{3,6}= 
    \tfrac{1}{120} e_{(5,2,2)} e_1^{3} 
    + \tfrac{1}{280} e_{(6,2)} e_1^{4} 
    - \tfrac{3}{140} e_{(6,3)} e_1^{3} 
    - \tfrac{229}{840} e_{(6,4)} e_1^{2}
    $ \\[.5ex]
	&    
    $\qquad \quad 
    - \tfrac{19}{40} e_{(6,5)} e_1 
    - \tfrac{1}{35} e_{(6,2,2)} e_1^{2} 
	+ \tfrac{141}{280} e_{(6,3,2)} e_1 
	+ \tfrac{17}{70} e_{(6,6)}
    $ \\[.5ex] 
	&
	$\qquad \quad
	+ \tfrac{3}{40} e_{(5,3,3)}e_1 
	- \tfrac{18}{35} e_{(6,3,3)}  
	- \tfrac{1}{10} e_{(5,3,2)} e_1^{2} 
	+ \tfrac{3}{5} e_{(6,4,2)}     
	$ \\[.5ex] 
	&
	$\qquad \quad
	+ \tfrac{1}{24} e_{(4,4,2)} e_1^{2} 
	+ \tfrac{1}{120} e_{(5,4)} e_1^{3} 
	+ \tfrac{19}{120} e_{(5,5)} e_1^{2} 
	- \tfrac{7}{40} e_{(5,4,2)} e_1 
	$ \\[0.5ex]
    $\bm{4}$ & $P_{4,6}=  
    - \tfrac{3}{280} e_{(6,3,2,2)} e_1^{2} 
    + \tfrac{9}{280} e_{(6,3,3,2)} e_1 
    + \tfrac{33}{280} e_{(6,6,3)}
    + \tfrac{51}{560} e_{(6,4,3,2)} 
    $ \\[.5ex]
	&
	$\qquad \quad    
	- \tfrac{9}{560} e_{(6,3,3,3)}  
    - \tfrac{5}{112} e_{(6,4,2,2)} e_1
    - \tfrac{3}{56} e_{(6,5,2,2)}  
    + \tfrac{1}{1680} e_{(6,2,2,2)} e_1^{3}
    $ \\[.5ex]
	&
	$ \qquad \quad
	+ \tfrac{11}{560} e_{(6,4,2)} e_1^{3}
    + \tfrac{29}{504} e_{(6,5,2)} e_1^{2}
    - \tfrac{1}{560} e_{(6,5)} e_1^{4} 
    - \tfrac{5}{126} e_{(6,4,4)} e_1
    $ \\[.5ex]
	&
	$\qquad \quad
	- \tfrac{1}{180} e_{(6,4,3)} e_1^{2} 
	+ \tfrac{11}{400} e_{(6,6,2)} e_1 
	- \tfrac{179}{5040} e_{(6,6)} e_1^{3}
    + \tfrac{1}{240} e_{(5,4,2,2)} e_1^{2} 
	$ \\[.5ex]
	&
	$\qquad \quad
	- \tfrac{1}{80} e_{(5,5,2)} e_1^{3} 
	+ \tfrac{1}{720} e_{(5,5,3)} e_1^{2}
    + \tfrac{1}{36} e_{(5,5,4)} e_1 
    + \tfrac{1}{80} e_{(5,5,2,2)} e_1 
	$ \\[.5ex]
	&
	$\qquad \quad	
	- \tfrac{1}{80} e_{(5,4,3,2)} e_1 
    - \tfrac{43}{840} e_{(6,5,3)} e_1 
    - \tfrac{5}{42} e_{(6,5,4)} 
	$ \\[.5ex]
    $\bm{5}$ & $P_{5,6}= 
    \tfrac{1}{3360} e_{(6,4,2,2,2)} e_1^{2} 
    - \tfrac{1}{1120} e_{(6,5,2,2)} e_1^{3} 
    - \tfrac{31}{7200} e_{(6,6,2,2)} e_1^{2}
    + \tfrac{9}{5600} e_{(6,6,2)} e_1^{4} 
	$ \\[.5ex]
	&
	$\qquad \quad
	+ \tfrac{13}{3360} e_{(6,5,3,2)} e_1^{2} 
	- \tfrac{281}{16800} e_{(6,6,4)} e_1^{2}
    + \tfrac{1}{800} e_{(6,6,3)} e_1^{3} 
    + \tfrac{121}{16800} e_{(6,6,5)} e_1 
    $ \\[.5ex]
	&
	$\qquad \quad
	- \tfrac{1}{3360} e_{(6,5,3,3)} e_1
    + \tfrac{11}{2100} e_{(6,6,4,2)}  
    + \tfrac{1}{4200} e_{(6,6,6)}  
    - \tfrac{1}{168} e_{(6,5,4,3)}  
	$ \\[.5ex]
	&
	$\qquad \quad
	+ \tfrac{1}{1260} e_{(6,5,2,2,2)} e_1 
	+ \tfrac{11}{4200} e_{(6,6,2,2,2)} 
	- \tfrac{1}{160} e_{(6,5,3,2,2)}
    - \tfrac{1}{360} e_{(6,4,3,2,2)} e_1  
	$ \\[.5ex]
	&
	$ \qquad \quad
	+ \tfrac{3}{1120} e_{(6,4,3,3,2)}  
	+ \tfrac{67}{5040} e_{(6,5,4,2)} e_1 
    + \tfrac{1}{280} e_{(6,5,5,2)}
    + \tfrac{1}{210} e_{(6,6,3,3)} 
	$ \\[.5ex]
	&
	$ \qquad \quad
	+ \tfrac{1}{336} e_{(6,4,4,2)} e_1^{2} 
    - \tfrac{1}{3360} e_{(6,5,4)} e_1^{3} 
    + \tfrac{1}{1440} e_{(5,5,3,2,2)} e_1 
    - \tfrac{1}{480} e_{(5,5,4,2)} e_1^{2} 
	$ \\[.5ex]
	&
	$\qquad \quad
	- \tfrac{1}{1440} e_{(5,5,5,2)} e_1 
	- \tfrac{127}{16800} e_{(6,6,3,2)} e_1 
	- \tfrac{1}{10080} e_{(6,5,5)} e_1^{2}
	$ \\[.5ex]
    $\bm{6}$ & $P_{6,6}= 
    \tfrac{1}{100800} e_{(6,6,2,2,2,2)} e_1 
    - \tfrac{1}{5600} e_{(6,6,3,2,2)} e_1^{2} 
    - \tfrac{11}{16800} e_{(6,6,4,2,2)} e_1
    $ \\[.5ex]
    &
    $ \qquad \quad
    + \tfrac{1}{1400} e_{(6,6,4,2)} e_1^{3}
    + \tfrac{1}{6720} e_{(6,6,3,3,2)} e_1 
    - \tfrac{1}{420} e_{(6,6,4,4)} e_1 
    $ \\[.5ex]
    &
    $\qquad \quad
    + \tfrac{1}{6720} e_{(6,6,4,3)} e_1^{2} 
    - \tfrac{97}{100800} e_{(6,6,6,2)} e_1 
    + \tfrac{59}{8400} e_{(6,6,6,3)}  
    $ \\[.5ex]
    &
    $ \qquad \quad
    + \tfrac{11}{25200} e_{(6,6,5,2,2)}  
    - \tfrac{49}{3600} e_{(6,6,5,4)}  
    + \tfrac{1}{4032} e_{(6,6,3,2,2,2)}  
    $ \\[.5ex]
    &
    $ \qquad \quad
	+ \tfrac{1}{20160} e_{(6,5,3,2,2,2)} e_1      
    - \tfrac{1}{6720} e_{(6,5,4,2,2)} e_1^{2} 
    + \tfrac{1}{120} e_{(6,5,5,5)}  
    $ \\[.5ex]
    &
    $ \qquad \quad
    - \tfrac{1}{3360} e_{(6,5,4,3,2)} e_1 
    - \tfrac{1}{6720} e_{(6,5,3,3,2,2)} 
    - \tfrac{47}{25200} e_{(6,6,5,3)} e_1 
	$ \\[.5ex]
	&
	$ \qquad \quad 
    + \tfrac{83}{33600} e_{(6,6,4,3,2)}  
    + \tfrac{1}{25200} e_{(6,6,5,2)} e_1^{2} 
    + \tfrac{1}{20160} e_{(6,5,5,3)} e_1^{2} 
    $ \\[.5ex]
    &
    $ \qquad \quad 
    + \tfrac{1}{2880} e_{(5,5,5,3,2)} e_1 
    - \tfrac{1}{2880} e_{(6,6,6)} e_1^{3} 
    - \tfrac{1}{10080} e_{(6,5,5,2,2)} e_1 
    $ \\[.5ex]
    &
    $ \qquad \quad 
    - \tfrac{13}{6720} e_{(6,5,5,3,2)}  
	+ \tfrac{1}{252} e_{(6,5,5,4)} e_1 
	- \tfrac{1}{720} e_{(5,5,5,5)} e_1
    $ \\[.5ex]
    $\bm{7}$ & $P_{7,6}= 
    \tfrac{1}{201600} e_{(6,6,3,3,2,2,2)}  
    - \tfrac{17}{201600} e_{(6,6,6,3,2)} e_1 
    + \tfrac{13}{25200} e_{(6,6,6,4,2)}   $ 
    \\[.5ex]
	&
	$\qquad \quad
	+ \tfrac{29}{16800} e_{(6,6,6,6)}
	- \tfrac{11}{28800} e_{(6,6,6,4)} e_1^{2} 
	+ \tfrac{1}{3360} e_{(6,6,6,3,3)}  
	$ \\[.5ex]
	&
	$\qquad \quad
	- \tfrac{11}{25200} e_{(6,6,5,4,3)}  
	+ \tfrac{1}{22400} e_{(6,6,4,4,2)} e_1^{2} 
	- \tfrac{1}{16800} e_{(6,6,4,3,2,2)} e_1 
	$ \\[.5ex]
	&
	$\qquad \quad	
	+ \tfrac{1}{11200} e_{(6,6,4,3,3,2)}  
	- \tfrac{1}{9600} e_{(6,6,5,3,2)} e_1^{2} 
	+ \tfrac{3}{22400} e_{(6,6,5,4,2)} e_1 
	$ \\[.5ex]
	&
	$\qquad \quad
	+ \tfrac{3}{5600} e_{(6,6,5,5)} e_1^{2} 
	+ \tfrac{1}{201600} e_{(6,6,5,2,2,2)} e_1 
	+ \tfrac{19}{201600} e_{(6,6,6,2,2)} e_1^{2} 
	$ \\[.5ex]
	&
	$\qquad \quad
	- \tfrac{1}{1400} e_{(6,6,5,5,2)}  
	+ \tfrac{1}{40320} e_{(6,5,5,3,2,2)} e_1 
	- \tfrac{1}{13440} e_{(6,5,5,3,3,2)}  
	$ \\[.5ex]
	&
	$\qquad \quad
	+ \tfrac{1}{3360} e_{(6,5,5,5,3)}  
	+ \tfrac{1}{6300} e_{(6,6,5,3,2,2)} 	
	- \tfrac{1}{25200} e_{(6,6,5,3,3)} e_1 
	$ \\[.5ex]
	&
	$\qquad \quad	
	- \tfrac{1}{16800} e_{(6,6,6,2,2,2)}  
	- \tfrac{1}{10080} e_{(6,5,5,5,2)} e_1 
	- \tfrac{11}{9600} e_{(6,6,6,5)} e_1
	$ \\[.5ex]
    $\bm{8}$ & $P_{8,6}= 
    \tfrac{1}{403200} e_{(6,6,5,3,3,2,2)}  
    - \tfrac{1}{134400} e_{(6,6,6,3,2,2,2)}  
    + \tfrac{1}{1209600} e_{(6,6,6,4,2,2)} e_1 
    $ \\[.5ex]
    &
    $ \qquad \quad
    + \tfrac{1}{403200} e_{(6,6,5,5,3,2)} 
    - \tfrac{1}{20160} e_{(6,6,5,5,5)}  
    + \tfrac{1}{33600} e_{(6,6,5,5,4)} e_1 
    $ \\[.5ex]
    &
    $ \qquad \quad
    + \tfrac{1}{134400} e_{(6,6,6,3,3,2)} e_1 
    - \tfrac{1}{134400} e_{(6,6,5,4,3,2)} e_1 
    + \tfrac{13}{67200} e_{(6,6,6,6,3)}  
	$ \\[.5ex]
	&
	$\qquad \quad    
    + \tfrac{1}{403200} e_{(6,6,6,3,3,3)}  
    + \tfrac{1}{75600} e_{(6,6,6,5,4)}  
    - \tfrac{31}{604800} e_{(6,6,6,5,3)} e_1 
    $ \\[.5ex]
	&    
    $\qquad \quad
    + \tfrac{1}{172800} e_{(6,6,6,4,3,2)}  
    + \tfrac{1}{60480} e_{(6,6,6,5,2)} e_1^{2} 
    - \tfrac{131}{1209600} e_{(6,6,6,6,2)} e_1 
    $ \\[.5ex]
    &
    $ \qquad \quad
	+ \tfrac{1}{28800} e_{(6,6,6,5,2,2)}  
    - \tfrac{1}{50400} e_{(6,6,6,4,4)} e_1
    $ \\[.5ex]
	$\bm{9}$ & $P_{9,6}= 
	\tfrac{1}{2419200} e_{(6,6,6,4,3,3,2)}  
	- \tfrac{1}{806400} e_{(6,6,6,5,3,2,2)}  
	- \tfrac{1}{2419200} e_{(6,6,6,6,3,2)} e_1 
	$ \\[.5ex]    
	&
	$\qquad \quad
	- \tfrac{1}{302400} e_{(6,6,6,6,4,2)}	
	+ \tfrac{1}{201600} e_{(6,6,6,5,5,2)}  
	+ \tfrac{1}{33600} e_{(6,6,6,6,6)}  
	$ \\[.5ex]
	&
	$ \qquad \quad
	+ \tfrac{1}{201600} e_{(6,6,6,6,3,3)}  
	- \tfrac{13}{604800} e_{(6,6,6,6,5)} e_1	
	- \tfrac{1}{604800} e_{(6,6,6,5,4,3)}
	$ \\[.5ex]
	$\bm{10}$ & $P_{10,6}= 
	\tfrac{1}{4838400} e_{(6,6,6,6,4,3,2)}  
	- \tfrac{1}{1209600} e_{(6,6,6,6,6,2)} e_1 
	+ \tfrac{1}{1612800} e_{(6,6,6,6,6,3)}  
	$ \\[.5ex]
    & $\qquad\quad
	- \tfrac{1}{1209600} e_{(6,6,6,6,5,4)}     
    $ \\[.5ex]
\bottomrule
\caption{The homogeneous components of $P_6$ in the basis of elementary symmetric polynomials.}
\label{table:hatP6:elementary}
\end{longtable}
\end{center}

\begin{center}
\begin{table}[H] 
\centering
\begin{tabular}[t]{r | c | c} 
\toprule
	\multicolumn{1}{c}{$\bm{(n,r)}$} & \multicolumn{1}{c}{\textbf{Appearing}}& \multicolumn{1}{c}{\textbf{Allowed}} \\ [.5ex]
\midrule
	$\bm{n=3}$, $\bm{r=0}$ & 1 & 1\\ [.5ex]
	$\bm{r=1}$ & 1 & 2\\ [.5ex]
\midrule
	$\bm{n=4}$, $\bm{r=0}$ & 1 & 1\\ [.5ex]
	$\bm{r=1}$ & 2 & 3\\ [.5ex]
	$\bm{r=2}$ & 2 & 5\\ [.5ex]
	$\bm{r=3}$ & 1 & 8\\ [.5ex]
\midrule
	$\bm{n=5}$, $\bm{r=0}$ & 1 & 1 \\ [.5ex]
	$\bm{r=1}$ & 3 & 4  \\ [.5ex]
	$\bm{r=2}$ & 5 & 8  \\ [.5ex]
	$\bm{r=3}$ & 7 & 13 \\ [.5ex]
	$\bm{r=4}$ & 6 & 20 \\ [.5ex]
	$\bm{r=5}$ & 3 & 28 \\ [.5ex]
	$\bm{r=6}$ & 2 & 38 \\ [.5ex]
\bottomrule
\end{tabular}
\begin{tabular}[t]{r | c | c} 
\toprule
    \multicolumn{1}{c}{$\bm{(n,r)}$} & \multicolumn{1}{c}{\textbf{Appearing}}& \multicolumn{1}{c}{\textbf{Allowed}} \\ [.5ex]
\midrule
    $\bm{n=6}$, $\bm{r=0}$ & 1 & 1 \\ [.5ex]
    $\bm{r=1}$ & 4 & 5 \\ [.5ex]
    $\bm{r=2}$ & 8 & 11 \\ [.5ex]
    $\bm{r=3}$ & 16 & 20 \\ [.5ex]
    $\bm{r=4}$ & 23 & 31 \\ [.5ex]
    $\bm{r=5}$ & 27 & 46 \\ [.5ex]
    $\bm{r=6}$ & 27 & 64 \\ [.5ex]
    $\bm{r=7}$ & 24 & 87 \\ [.5ex]
    $\bm{r=8}$ & 17 & 114 \\ [.5ex]
    $\bm{r=9}$ & 9 & 148 \\ [.5ex]
    $\bm{r=10}$ &4  & 187 \\ [.5ex]
\bottomrule
\end{tabular}
\caption{$P_{r,n}$ in the basis of elementary symmetric polynomials: number of partitions \textbf{appearing} (i.e allowed with non-zero coefficients) vs number of partitions \textbf{allowed} (length$\leq r$ and weight $\leq d_{r,n}$). }
\label{table:AllowedvsAppearingCoeff}
\end{table}
\end{center}

\printbibliography
\end{document}